\documentstyle[12pt]{article}
\input epsf.tex

\newcommand{\bmat}{\left(\begin{array}}
\newcommand{\emat}{\end{array}\right)}

\def\preal{{\rm Re\,}}
\def\pim{{\rm Im\,}}

\def\yzero{\smash{\hbox{$y\kern-4pt\raise1pt\hbox{${}^\circ$}$}}}

\def\-{\hphantom{-}}
\def\ov{\overline}
\def\s2{\frac{1}{\sqrt2}}

\def\beq{\begin{equation}}
\def\eeq{\end{equation}}
\def\beqa{\begin{eqnarray}}
\def\eeqa{\end{eqnarray}}

\def\Tr{{\rm Tr \,}}
\def\diag{{\rm diag \,}}
\def\IC{\relax\hbox{\kern.25em$\inbar\kern-.3em{\rm C}$}}

\def\IC{\bf C}

\def\IZ{\bf Z}
\def\IP{\bf P}

\def\Dsl{\,\raise.15ex\hbox{/}\mkern-13.5mu D} 
\def\id{{\rm I}}

\catcode`\@=11
\newdimen\@rotdimen
\newbox\@rotbox

\def\@vspec#1{\special{ps:#1}}
\def\@rotstart#1{\@vspec{gsave currentpoint currentpoint translate
   #1 neg exch neg exch translate}}
\def\@rotfinish{\@vspec{currentpoint grestore moveto}}
\def\@rotr#1{\@rotdimen=\ht#1\advance\@rotdimen by\dp#1%
   \hbox to\@rotdimen{\hskip\ht#1\vbox to\wd#1{\@rotstart{90 rotate}%
   \box#1\vss}\hss}\@rotfinish}
\def\@rotl#1{\@rotdimen=\ht#1\advance\@rotdimen by\dp#1%
   \hbox to\@rotdimen{\vbox to\wd#1{\vskip\wd#1\@rotstart{270 rotate}%
   \box#1\vss}\hss}\@rotfinish}%
\def\@rotu#1{\@rotdimen=\ht#1\advance\@rotdimen by\dp#1%
   \hbox to\wd#1{\hskip\wd#1\vbox to\@rotdimen{\vskip\@rotdimen
   \@rotstart{-1 dup scale}\box#1\vss}\hss}\@rotfinish}%
\def\@rotf#1{\hbox to\wd#1{\hskip\wd#1\@rotstart{-1 1 scale}%
   \box#1\hss}\@rotfinish}%
\def\rotate{\@ifnextchar[{\@rotate}{\@rotate[l]}}
\def\@rotate[#1]#2{\setbox\@rotbox=\hbox{#2}\@nameuse{@rot#1}\@rotbox}

\catcode`\@=12

\topmargin
-1.5cm
\textwidth
15.5cm
\textheight
23.5cm
\oddsidemargin
0.7cm
\evensidemargin
1.2cm

\begin{document}

\makeatletter
\@addtoreset{equation}{section}
\makeatother
\renewcommand{\theequation}{\thesection.\arabic{equation}}
\pagestyle{empty}
\rightline{ IFT-UAM/CSIC-04-36}
\rightline{\tt hep-th/yymmnn}
\vspace{0.5cm}
\begin{center}
\LARGE{Flux-induced SUSY-breaking soft terms
on D7-D3 brane systems  \\[10mm]}
\large{
P. G. C\'amara$^*$, L.~E.~Ib\'a\~nez$^{*,**}$ and A. M. Uranga$^*$
\\[2mm]}
\small{
*\ Departamento de F\'{\i}sica Te\'orica C-XI
and Instituto de F\'{\i}sica Te\'orica  C-XVI,\\[-0.3em]
Universidad Aut\'onoma de Madrid,
Cantoblanco, 28049 Madrid, Spain.\\ [4mm]
**\ Theory Division, CERN, 1211 Geneva 23, Switzerland.
\\[9mm]}
\small{\bf Abstract} \\[7mm]
\end{center}

\begin{center}
\begin{minipage}[h]{14.0cm}
{\small
We study the effect of RR and NSNS 3-form fluxes on the effective action
of the worldvolume fields of Type IIB D7/D3-brane  configurations. The
D7-branes wrap 4-cycles $\Sigma_4$ on a local Calabi-Yau geometry.
This is an extension of previous work on hep-th/0311241, where a
similar analysis was applied to the case of D3-branes. Our present
analysis is based on the  D7- and D3-brane Dirac-Born-Infeld and
Chern-Simons actions, and makes full use of the R-symmetries of the
system, which allow us to compute explicitly results for the fields lying
at the D3-D7 intersections. A number of interesting new properties appear
as compared to the simpler case of configurations with only D3-branes. As
a general result one finds that fluxes stabilize some or all of the
D7-brane moduli. We argue that this is important for the
problem of stabilizing K\"ahler moduli through non-perturbative effects in
KKLT-like vacua. 
 We also show that (0,3) imaginary self-dual fluxes,
which lead to compactifications with zero vacuum energy,  give rise to
SUSY-breaking soft terms including gaugino and scalar masses, and
trilinear terms. Particular examples of chiral MSSM-like models of this class of
vacua, based on D3-D7 brane systems at orbifold singularities are
presented.
}

\end{minipage}
\end{center}
\newpage
\setcounter{page}{1}
\pagestyle{plain}
\renewcommand{\thefootnote}{\arabic{footnote}}
\setcounter{footnote}{0}


\section{Introduction}

In the search for realistic string compactifications two important 
problems appear. On one hand, in e.g. familiar Calabi-Yau 
compactifications, one typically has a large number of 
perturbatively massless fields, namely the complex dilaton, K\"ahler 
and complex structure moduli, and possibly others (e.g. gauge bundle or 
brane moduli). Thus the string coupling as well as geometric data of the 
compactification remain undetermined at the perturbative level. A second 
problem, possibly related to the first one, is how to break supersymmetry 
in a controled manner if we start with e.g. a $N=1$ supersymmetric 
compactification.

In the last few years it has been realized that a general compactification 
allows for an additional ingredient, not considered previously, namely 
field strength fluxes for the internal components of $p$-form supergravity 
fields 
\cite{earlyflux, beckers, gvw,drs,superp,gkp,moreflux,fluxmodels,ferrara}. 
Interestingly, this new ingredient may have an important bearing on those
two problems. The case of NSNS/RR 3-form field strength fluxes in Type IIB 
CY orientifold compactifications (or more generally, F/M-theory on CY 
fourfolds with 4-form flux) has been studied in particular detail (see 
e.g. \cite{beckers, drs,gkp}). The presence of the fluxes induces a 
non-trivial warp factor in the geometry, as well as a non-trivial 
source for the RR 4-form potential. However it has been shown that, due to 
the presence of O3-planes in the IIB construction (or more generally, 
the F/M-theory tadpole from the Euler characteristic of the fourfold 
\cite{svw}), the Type IIB 10-dimensional equations of motion can be 
solved consistently with 4d Minkowski space, if the internal 3-form flux 
is imaginary self-dual, ISD (analogously, if the F/M-theory 4-form flux 
is self-dual). 

The fact that the (quantised) fluxes should be ISD to solve the equations 
of motion generically determines dynamically the value of the complex 
dilaton $\tau$ as well as all complex structure moduli, while K\"ahler 
moduli are not stabilized. For adequate choices of fluxes, the complex 
dilaton may be fixed at a perturbative value, so that one expects the 
description of the compactifications in terms of classical supergravity to
be a reliable approximation (in the large volume limit in K\"ahler moduli 
space). It has further been argued that non-perturbative effects depending 
on the K\"ahler moduli, which are generically present in this kind of 
compactification, like non-perturbative superpotentials from 
strong infrared dynamics of gauge sectors on the D7-branes or from
euclidean D3-brane instantons, may determine also all the K\"ahler moduli
\cite{kklt} (possibly at large volume). This has been further explored in 
explicit models in \cite{ddf,sethi}, with positive results in a sizeable 
number of examples \cite{ddf}. This setting thus provides the first 
examples of tractable string compactifications with all dilaton, K\"ahler 
and complex structure moduli determined dynamically.

Field strength fluxes may also be important for the second problem 
mentioned above, the breaking of supersymmetry, since the generic choice 
of fluxes in a compactification is in fact non-supersymmetric (although 
many examples with supersymmetry preserving fluxes have been constructed).
In particular the field theories inside D$p$-branes located in the CY 
compactification, which are supersymmetric in the absence of fluxes, may 
get SUSY-breaking soft terms in the presence of fluxes. This is 
particularly interesting since it is possible to embed chiral gauge 
sectors relatively close to the (MS)SM in D-brane configurations in the 
presence of 3-form fluxes \cite{blt,cascur,msw}. In the case of 
gauge sectors localized on D3- (or anti-D3-) branes, such terms have been
recently computed in \cite{grana,ciu,ggjl}. In particular it was found in 
\cite{ciu} (see also \cite{ggjl}) that only imaginary anti-selfdual (IASD) 
fluxes gives rise to SUSY-breaking soft terms on the worldvolume of 
D3-branes \footnote{On the other hand ISD do give rise to soft terms on
the worldvolume of anti-D3-branes. The presence of these however
break supersymmetry in a less controlled fashion.}. This is unfortunate 
because, as we mentioned, only ISD fluxes solve the equations of motion. 
Thus, in order to have non-vanishing soft-terms,  we either add IASD 
fluxes, and hence do not solve the equations of motion at this level 
(hoping that other effects would perhaps stabilize the compactification), 
or else we include anti-D3-branes in the compactification, thus 
breaking supersymmetry in a less controlled fashion. The latter is 
certainly an interesting possibility, as advocated in \cite{ciu}, since
anti-D3-branes turn out to be an important ingredient in the proposal in
\cite{kklt} to obtain deSitter vacua in string theory.

In any event, it would be interesting to have Type IIB CY orientifold 
compactifications (or more generally F-theory compactifications on CY 
4-folds) which have $N=1$ supersymmetry in the absence of fluxes, and
yield SUSY-breaking soft terms upon turning them on, and still obeying the
Type IIB equations of motion. One of the motivations of the present paper 
is to provide examples of this class. Specifically we analyze the effect 
of 3-form field strength fluxes on the world-volume fields of D7/D3-brane 
configurations, extending the results for D3-branes presented in 
\cite{ciu,ggjl}. We show that, in contrast with the D3-brane case, ISD 
fluxes do give rise to non-vanishing soft terms for the fields  
on the worldvolume of D7-branes.

This result is interesting for a number of reasons. In particular this 
shows that Type IIB CY orientifolds in the presence of ISD fluxes provide 
us with (to our knowledge) the first known class of string compactifications
to 4d Minkowski space which solve the classical equations of motion and
lead to non-trivial SUSY-breaking soft terms, in a controled manner. 
Apart from its theoretical interest, the structure of soft terms, if
applied to realistic models with the SM living on D7-branes,
may be of phenomenological relevance (see \cite{fluxedmssm}).

Another application of our results concerns the proposal in \cite{kklt}, 
attempting to fix dynamically the K\"ahler moduli in type IIB orientifolds 
of this class, mentioned above. A potential problem in generating 
non-perturbative superpotentials from e.g. strong infrared dynamics on the 
gauge theory on D7-branes, is the possible presence of too much 
massless charged matter in the latter. Our results for soft terms
involving D7-brane matter fields show that the presence of ISD fluxes 
generally give masses to all the D7-brane geometric moduli. This indicates 
that generically D7-brane vector-like matter is massive, thus allowing
such non-perturbative superpotentials to appear \footnote{An important 
exception is D7-branes containing charged chiral fermions. This suggests 
that the D7-branes responsible for K\"ahler moduli stabilization should 
not correspond to the D7-branes on which we plan to embed the SM,
but to some non-chiral D7-brane sector.}. This 
is also in agreement with recent results derived from the F-theory 
perspective in \cite{gktt}, and from generalized calibrations 
\cite{calgen}. 

Our results are based on an expansion of the relevant D7- and D3-brane
Dirac-Born-Infeld plus Chern-Simons (DBI+CS) actions in the presence of 
fairly general type IIB closed string backgrounds. Since D-branes are 
mainly sensitive to the local geometry around them, our results 
are derived in an expansion around the location of D7- and D3-branes. 
Since D7-branes wrap a 4-cycle $\Sigma_4$ in the internal space, we center 
on the simple situation of D7-branes wrapping $T^4$ or $K3$ in a local 
Calabi-Yau
$\Sigma_4\times C$. The effect of fluxes on the fields living at D3-D7 
(or ${\ov {D3}}-D7$) intersections is more involved, but we explicitly 
derive it in full generality using (super)symmetry arguments.
Finally, all results are compared, showing full agreement, with an 
analysis based on the use of the 4d effective action \cite{imr}.

The paper is organized as follows. After reviewing the results of 
\cite{ciu} in section 2, we compute the soft terms for the fields on the
worldvolume of D7's in section 3, checking also that our ansatz for the
closed string backgrounds solves the equations of motion. In sections 
4 (resp. 6) we make use of the local geometric symmetries of $D3$-$D7$ 
(resp. ${\ov {D3}}-D7$) configurations to carry out the computation
of the soft terms for  fields living at the $D3$-$D7$( ${\ov {D3}}$-$D7$)  
intersections. Examples of soft terms obtained for different classes of 
ISD and IASD fluxes are described in section 5.

Some of the results obtained both for $D7-D3$-brane systems may be 
understood from the effective low energy $N=1$ supergravity action
\cite{imr,lrs,louis}. In 
particular, we show in section 7 that fluxes correspond to non-vanishing
expectation values for the auxiliary fields of the dilaton and K\"ahler 
moduli. We show that the soft terms obtained in section 3  for ISD 
$(0,3)$ fluxes may be understood as arising from a non-vanishing auxiliary 
field $F_T\not=0$ for the overall K\"ahler modulus $T$. This is in 
agreement with previous results in \cite{ciu,ggjl}. A noteworthy
property is that the bosonic soft terms induced by ISD fluxes 
naturally combine with the SUSY scalar potential into a positive definite 
scalar potential. This fact, which appears already in section 3 from
the  DBI+CS action in the presence of fluxes, has a simple interpretation 
also from the effective $N=1$ supergravity low-energy action, as explained
in section 7. We also show in that section that analogous soft terms are 
expected for the fields lying at the intersections of $D7$-branes   
wrapping different 4-cycles.

In section 8 we argue that  many of the results obtained for D7-branes 
wrapping 4-tori are still valid when D7-branes wrap curved 4-cycles, in 
particular $K3$, and describe the explicit soft terms in this case.
We also argue that similar patterns of soft terms are also obtained in more 
involved situations, where the normal bundle to the 4-cycle is 
non-trivial. For instance, in cases with multiple D7-brane geometric 
moduli, all of the latter are fixed in the presence of ISD backgrounds,
in agreement with results in section 7.

Some particular examples and applications are described in section 9. In 
particular we present several examples of local D7/D3-brane  
configurations, with a semirealistic chiral gauge sector arising in the 
worldvolume of the D7-branes. The presence of ISD fluxes gives rise to 
phenomenologically interesting soft terms. Embedding this class of local 
configurations into a full F-theory compactification would give rise to 
semirealistic models in which calculable SUSY-breaking soft terms are 
induced. We also briefly discuss in section 9 the relevance of our results 
for the program in \cite{kklt}, as discussed above. We end up with some 
comments in section 10. Details of the computations are provided in an 
appendix.

\medskip

\section{Three-form fluxes and  D3-branes}
\label{dthree}

Let us first briefly review some of the main results in ref.\cite{ciu} 
(see \cite{grana,ggjl} for related discussions).
We consider D3-branes embedded in general Type IIB closed string 
backgrounds of the general form
\beqa
ds^2 & = & Z_1(x^m)^{-1/2} \eta_{\mu\nu}\, dx^\mu dx^\nu + Z_2(x^m)^{1/2}
\, ds^2{}_{CY} \nonumber \\
\tau & = & \tau(x^m) \nonumber \\
G_3 & = & \frac{1}{3!}G_{lmn} (x^m)\, dx^l dx^m dx^n \nonumber \\
\chi_4 & = & \chi(x^m) \, dx^0 dx^1 dx^2 dx^3 \\
F_5 & = & d\chi_4 + *_{10}\, d\chi_4  \nonumber
\label{ansatz}
\eeqa
where $G_3=F_3-\tau H_3$ (with $F_3(H_3)$  being the RR(NSNS) flux), and
$ds^2{}_{CY}$ denotes the metric in transverse space, in the absence of
flux backreaction, i.e. the Ricci-flat metric of the underlying
Calabi-Yau. Eventually we will request these backgrounds to solve  the
Type IIB supergravity equations of motion. Since D3-branes are located 
at a point in the six dimensions parametrized by $x^m$, it is natural to 
expand  these backgrounds around the position $x^m$ of the D3-branes
\beqa
Z_1^{-1/2} & = & 1 + \frac{1}{2} K_{mn}\, x^m x^n + \ldots \nonumber \\
Z_2^{1/2} & = & 1 + \ldots \nonumber\\
\tau & = &  \tau_0 + \frac{1}{2}\tau_{mn}\, x^m x^n \nonumber \\
\chi_4 & = & ({\rm const.} +\frac{1}{2}\chi_{mn}\, x^mx^n+\ldots) \,
dx^0 dx^1 dx^2 dx^3 \\
G_{lmn} (x^m)  & = & G_{lmn} + \ldots \nonumber
\label{powerexp1}
\eeqa
where the coefficients $K$, $F$, $G$, $\tau$ in the right hand side are
constant, independent of $x^m$. The piece of the 5-form background
relevant for our purposes below is
\beqa
F_5 & = &
 \frac{1}{2}(\chi_{mn}+\chi_{nm}) x^m dx^n dx^0 dx^1 dx^2 dx^3
+\ldots
\label{powerexp2}
\eeqa

In the absence of fluxes, the massless fields on a stack of $n$
D3-branes are given by $U(n)$ gauge bosons, six real adjoint scalars, and
four Majorana adjoint fermions, the latter transforming in the
representations ${\bf 6}$ and ${\bf\ov 4}$ of the $SO(6)$ local symmetry. 
In $N=1$
supersymmetry language, they fill out a vector multiplet and three chiral
multiplets. The effect of the
above backgrounds on the effective action of the worldvolume fields can be
obtained by plugging the above expansions in the DBI and CS action for
the D3-branes.

Calabi-Yau manifolds are endowed with a complex structure, hence
it is useful to rewrite the world-volume fields and the 3-form background
in complex coordinates. Let us introduce local complex coordinates
$z^1=\frac{1}{\sqrt{2}}(x^4+ix^5)$, $z^2=\frac{1}{\sqrt{2}}(x^6+ix^7)$,
$z^3=\frac{1}{\sqrt{2}}(x^8+ix^9)$. We define the complex scalars
$\Phi^1=\frac{1}{\sqrt{2}}(\phi^4+i\phi^5)$,
$\Phi^2=\frac{1}{\sqrt{2}}(\phi^6+i\phi^7)$, $\Phi^3=\frac{1}{\sqrt{2}}
(\phi^8+i\phi^9)$, and use $\Phi^{\bar{i}}$ to denote $(\Phi^i)^*$.
Denoting the four-plet of fermions by their $SO(6)$ weights, the fermion
$\frac 12(+++)$ belongs to the $N=1$ vector multiplet, and is referred to
as the gaugino, denoted $\lambda$. The fermions combining with the
above complex scalars to give $N=1$ chiral supermultiplets are $\Psi^1$,
$\Psi^2$, $\Psi^3$, corresponding to the weights $\frac 12(+--)$, $\frac
12(-+-)$ and $\frac 12 (--+)$, respectively.

In a preferred complex structure, only an $SU(3)$ (times $U(1)$) subgroup
of the $SO(6)$ symmetry is manifest, hence it is useful to decompose the 
3-form flux background under it. The antisymmetric flux $G_{mnp}$
transforms as a 20-dimensional reducible $SO(6)$ representation,
decomposing as ${\bf 20}={\bf{\ov{10}}}+{\bf 10}$. The irreducible
representations
${\bf {\ov{10}}}$, ${\bf 10}$ correspond to the imaginary self-dual (ISD)
$G_{(3)}^+$
and imaginary anti self-dual (IASD) $G_{(3)}^-$ parts, respectively,
defined as
\beq
G_{(3)}^{\pm}\ =\ {1\over 2} (G_{(3)}\mp i*_6 G_{(3)})
\ \ ;\ \  *_6G_{(3)}^{\pm}\ =\ \pm iG_{(3)}^{\pm}
\label{ISD}
\eeq
It is useful to classify the
components of the ISD and IASD parts of $G_3$  according to their
behaviour under $SU(3)$, ${\bf 10}={\bf 6}+{\bf 3}+{\bf 1}$. For that
purpose, we introduce the tensors \cite{gp}
\beqa
S_{ij} &=&
\frac{1}{2}(\epsilon_{ikl}G_{j\bar{k}\bar{l}}+
\epsilon_{jkl}G_{i\bar{k}\bar{l}}) \nonumber\\
A_{\bar{i}\bar{j}} &=&
\frac{1}{2}(\epsilon_{\bar{i}\bar{k}\bar{l}}G_{kl\bar{j}}-
\epsilon_{\bar{j}\bar{k}\bar{l}}G_{kl\bar{i}}) \nonumber
\label{sa}
\eeqa
defined in terms of the complex components of $G_3$, and which transform
in the representation ${\bf 6}$ and ${\bf 3}$ under $SU(3)$.
One similarly defines $S_{\bar{i}\bar{\j}}$ and $A_{ij}$.
The $SU(3)$ properties of each component is displayed in table 1.

\begin{table}[htb]
\renewcommand{\arraystretch}{0.70}
\begin{center}
\begin{tabular}{|c|c|c||c|c|c|}
\hline      & ISD   &     &   & IASD &   \\
\hline   $SU(3)$ rep. &  Form  &  Tensor &  $SU(3)$ rep. & Form & Tensor
\\
\hline \hline
 ${\overline 1}$  & $(0,3)$  &  $G_{{\bar 1}{\bar 2}{\bar 3}}$ &
 $ 1$  & $(3,0)$  &  $G_{{ 1}{ 2}{ 3}}$ \\
\hline
 ${\overline 6}$  & $(2,1)_P$  &  $ S_{{\bar i}{\bar j}}$ &
 ${ 6}$  & $(1,2)_P$  &  $ S_{{ i}{ j}}$ \\
\hline
${\overline 3}$  & $(1,2)_{NP}$  &  $ A_{{ i}{ j}}$ &
 ${ 3}$  & $(2,1)_{NP}$  &  $ A_{{\bar  i}{\bar j}}$ \\
\hline \end{tabular}
\caption{$SU(3)$ decomposition of antisymmetric $G_{(3)}$ fluxes.}
\end{center}
\end{table}

In \cite{ciu} the soft term Lagrangian for
the worldvolume fields of D3-branes was computed to be
\beqa
{\cal L}  =  \Tr \biggr[&
-\, (\,2K_{i\bar{\j}}-\chi_{i\bar{\j}}+g_s (\pim \tau)_{i\bar{\j}}\, )\,
\Phi^i\Phi^{\bar{\j}}\, -\, \frac{1}{2}\,(\, 2K_{ij}-\chi_{ij}+g_s(\pim
\tau)_{ij}\, )\, \Phi^i\Phi^j\, +\, {\rm h.c.}  \nonumber \\
& +g_s\sqrt{2\pi}\, \left[\, \frac{1}{3}G_{123}\, \epsilon_{ijk}\,
\Phi^i\Phi^j\Phi^k + \frac{1}{2}\epsilon_{\bar{i}\bar{j}\bar{l}}\,
(S_{lk}-(A_{\bar{l}\bar{k}})^*)\, \Phi^{\bar{i}}\Phi^{\bar{j}}\Phi^k\, +\,
{\rm h.c.} \, \right]\, + \nonumber \\
&+\, \frac{g_s^{1/2}} {2\sqrt{2}} \, \left[ \, G_{123}\, \lambda\lambda
\, +\, \frac{1}{2}\epsilon_{ijk}A_{\bar{j}\bar{k}}\, \Psi^i\lambda \, +\,
\frac{1}{2}S_{ij}\, \Psi^i\Psi^j \, +\, {\rm h.c.}\, \right]\,\, \biggr]
\label{softgencomp}
\eeqa
where we have defined
\beq
(\pim \tau)_{i\bar{\j}} =
\frac{1}{2i}(\tau_{i\bar{\j}}-(\tau_{j\bar{i}})^*) \ \ ;\ \
(\pim\tau)_{ij} = \frac{1}{2i}(\tau_{ij}-(\tau_{\bar{\j}\bar{i}})^*)
\eeq
In the notation of the first appendix in \cite{ciu} one thus has soft
terms for the D3-brane worldvolume fields
\beqa
m_{ij}^2\ & =&\ 2K_{i\bar{\j}}-\chi_{i\bar{\j}}+g_s (\pim \tau)_{i\bar{\j}}
\nonumber
\\
B_{ij}\ &=& \ 2K_{ij}-\chi_{ij}+g_s(\pim \tau)_{ij} \nonumber
\\
A^{ijk}\ &=& \ -h^{ijk} {{g_s^{1/2}}\over {\sqrt{2}}}\ G_{123}
 \nonumber
\\
C^{ijk}\ &=& +h^{ijl} {{g_s^{1/2}}\over {2\sqrt{2}}}(S_{lk}-(A_{{\bar l}{\bar
k}})^*)
\nonumber
\\
M^a\ &=& \  {{g_s^{1/2}}\over {\sqrt{2}}}\ G_{123} \nonumber
\\
\mu _{ij} \ & =& \  -{{g_s^{1/2}}\over {2\sqrt{2}}} S_{ij}
 \nonumber
\\
M_{g}^{ia}\ &=&\ {{g_s^{1/2}}\over {4\sqrt{2}}}
\epsilon_{ijk} A_{{\bar j}{\bar k}}
\label{softgen}
\eeqa
The supergravity equations of motion impose some constraints on the
background, namely \cite{ciu}
\begin{equation}
4\sum K_{l\bar{l}}\, = \,\frac{g_s}{2}\left(\vert G_{123}\vert^2
+ \vert G_{\bar{1}\bar{2}\bar{3}}\vert^2 +
\frac{1}{4}\sum_{ij}(\vert S_{ij}\vert^2 +
\vert S_{\bar{i}\bar{j}}\vert^2 + \vert A_{ij}\vert^2
+ \vert A_{\bar{i}\bar{j}}\vert^2)\right)
\label{tracemetric}
\end{equation}
\begin{equation}
-2\sum \chi _{l\bar{l}}\, =\, \frac{g_s}{2}\, \left(\, \vert
G_{123}\vert^2
\, -\, \vert G_{\bar{1}\bar{2}\bar{3}}\vert^2\, +\,
\frac{1}{4}\, \sum_{ij}(\, \vert S_{ij}\vert^2 \, -\,
\vert S_{\bar{i}\bar{j}}\vert^2\, -\,
\vert A_{ij}\vert^2\, +\, \vert A_{\bar{i}\bar{j}}\vert^2\,)\, \right)
\quad 
\end{equation}
\beqa
i\sum\tau_{l\bar{l}}\, =\, \frac{1}{2}\, \left(\,
G_{123}G_{\bar{1}\bar{2}\bar{3}}\, +\, \frac{1}{4}\,
S_{lk}S_{\bar{l}\bar{k}}\, +\, \frac{1}{4}\, A_{lk}A_{\bar{l}\bar{k}}\,
\right) 
\eeqa
They allow to relate the scalar masses to the flux background, as follows
\beqa
m_1^2 \, +\,  m_2^2\,  +\, m_3^2\, =\,
\frac{g_s}{2}\, \left[ \, \vert G_{123} \vert^2\, +\,
\frac{1}{4}\, \sum_{ij}\, (\, \vert S_{ij} \vert^2\, +\,
\vert A_{\bar{i}\bar{j}}\vert^2\, )\, -\, \nonumber \right. \\
\left. -\, \preal\, (\, G_{123}G_{\bar{1}\bar{2}\bar{3}}\, +\,
\frac{1}{4}\, S_{lk}S_{\bar{l}\bar{k}}\, +\,
\frac{1}{4}\, A_{lk}A_{\bar{l}\bar{k}})\, \right]
\label{genmass}
\eeqa
As emphasized in \cite{ciu}, the scalar mass matrix is not fully
determined in terms of the fluxes, only its trace is.

Our purpose in the present paper is to carry out a similar analysis for
the effect of fluxes on configurations including D7-branes.

\section{D7-branes and RR, NSNS 3-form fluxes}
\label{dseven}

\subsection{D7-branes and the 3-form flux background}

There are a number of differences between the case of D7 and that of
D3-branes reviewed in the previous section. First of all the D7-branes
in general wrap a 4-cycle $\Sigma_4$ on the internal space. Thus an
expansion of the closed string background on all 6 transverse coordinates
$x^m$  as we did in the D3-brane case is no longer the correct
procedure. Rather, we need to describe the local geometry around the 
4-cycle wrapped by the D7-brane, namely a tubular neighbourhood around the 
4-cycle, given by the normal bundle of the 4-cycle, i.e. the wrapped 
4-cycle, on which we fiber the normal direction. Then we may expand the 
background in a power series in
its dependence on the normal coordinate (while in principle keeping its
full dependence on $\Sigma_4$). For most purposes, we will center on
cases where the fibration is trivial, and the local geometry is 
$\Sigma_4\times C$. \footnote{This restricts us to backgrounds
$T^4\times C$ or K3$\times C$, which is a restricted set, but sufficiently
rich to have non-trivial effects. Moreover, the qualitative features can
be extrapolated to more involved situations, see section 8.} 
We will comment later on possible generalizations of this background.

The strategy then is to obtain the 8d action for the D7-brane,
taking the effect of fluxes into account. An important ingredient
is that the symmetry of the problem is reduced as compared with the 
D3-brane case. In fact, the
supergravity background need only respect 4d Lorentz invariance.
On the other hand, the local geometric symmetry in the internal
space is only $SO(4)\times SO(2)$, where the first factor is the
local euclidean rotation group on $\Sigma_4$, and the second in the
normal direction. Hence the local 8d flux-induced terms must
respect that symmetry (taking into account its action on the
world-volume fields as well).

A second step involves the Kaluza-Klein compactification
of this 8d action on $\Sigma_4$ to 4d. In this process, the geometry of
$\Sigma_4$ plays a crucial role even in the absence of fluxes, and
determines the number and kind of the massless fields on the D7-brane
gauge sector in 4d. Our purpose is to determine the effect of the 8d flux
induced terms on these massless fields, hence to compute it we will need
to center on concrete cases. Our analysis will mainly center on the case
of $T^4$ (which can be subsequently employed to study orbifolds
thereof), and K3, with trivial worldvolume gauge bundles.

We are thus interested in determining first the 8d action,
including the effects of fluxes, and second in discussing the
compactification to 4d. We will concentrate on the most relevant
terms, namely up to dimension three, hence scalar masses, scalar
trilinears and fermion masses.

\medskip

Still this is a rather involved problem in the general case, and we
will have to make some restricting assumptions on the geometry. In
particular, we assume certain restricted classes of closed string
backgrounds. The bottomline is that we consider the flux to have the 
structure in eq. (3.6), and to be pure ISD or pure IASD; we will 
also center in situations where the background is constant over $\Sigma_4$.

To specify our assumptions and also for practical purposes it is now 
useful to write the 3-form fluxes $G_{mnp}$ in terms of the local 
geometric symmetry $SO(4)\times SO(2)$. Thus now the decomposition of the 
ISD (IASD) pieces of $G_3$ in representations of the $SU(2)\times 
SU(2)'\times U(1)$ symmetry are as follows
\beq
10 \ =\ (3,1)_-+(1,3')_++(2,2')_0 \ \  ,\ \
\overline{10}\ =\ (3,1)_++(1,3')_-+(2,2')_0
\eeq
where the subindex gives the $\pm 1$ $U(1)$ charge. Choosing our 4-cycle
to be parametrized by $z^1$, $z^2$, and localized in the transverse
direction $z^3$, the triplets of $SU(2), SU(2)'$
are related with the fluxes in $SU(3)$ notation by
\beqa
(1,3')_+&=&\{{\mathcal{G}}_0' = -\frac{1}{\sqrt{2}}
A_{\bar{1}\bar{2}}\ ,\
{\mathcal{G}}'_x = -\frac{1}{\sqrt{2}}(\frac{1}{2}S_{33}-G_{123})\ ,\
{\mathcal{G}}'_y = -\frac{i}{\sqrt{2}}(\frac{1}{2}S_{33}+G_{123})
\} \nonumber \\
(3,1)_-&=&\{ {\mathcal{G}}_0 =  \frac{1}{\sqrt{2}}S_{12}\ ,\
{\mathcal{G}}_x = -\frac{1}{\sqrt{2}}(\frac{1}{2}S_{11}-\frac{1}{2}S_{22}) \ ,\
 {\mathcal{G}}_y = -\frac{i}{\sqrt{2}} 
(\frac{1}{2}S_{11}+\frac{1}{2}S_{22})
 \}\nonumber\\
(1,3')_-&=&\{G_0' =  -\frac{1}{\sqrt{2}}A_{12}\ ,\
G'_x = -\frac{1}{\sqrt{2}}(\frac{1}{2}S_{\bar{3}\bar{3}}-G_{\bar{1}\bar{2}\bar{3}}) \ ,\
G'_y = -\frac{i}{\sqrt{2}}(\frac{1}{2}S_{\bar{3}\bar{3}}+G_{\bar{1}\bar{2}\bar{3}})
\}\nonumber \\
(3,1)_+&=&\{G_0 =  \frac{1}{\sqrt{2}}S_{\bar{1}\bar{2}}\ ,\
G_x = -\frac{1}{\sqrt{2}}(\frac{1}{2}S_{\bar{1}\bar{1}}-\frac{1}{2}S_{\bar{2}\bar{2}}) \ ,\
 G_y = -\frac{i}{2}(\frac{1}{2}S_{\bar{1}\bar{1}}+\frac{1}{2}S_{\bar{2}\bar{2}})
\}
\label{flujosso4}
\eeqa
Regarding the triplets of $SU(2)$ groups as vectors of $SO(3)$, for future 
convenience we define the $SU(2)$ invariant scalar product
\beq
A\cdot B = A_0B_0+A_xB_x+A_yB_y 
\label{product}
\eeq
On the other hand the $G_{mnp}$ components transforming like $(2,2)_0$
correspond to the SU(3) components $S_{i3}, A_{i3}, S_{\bar{i}\bar{3}},
A_{\bar{i}\bar{3}}, i=1,2$. These fluxes are special in several respects.
In particular, if $\Sigma_4$ contains 3-cycles $C_3$ , this multiplet
contains fluxes such that
\beq \int_{C_3} F_3\neq 0\quad ; \quad \int_{C_3} H_3\neq 0 \eeq
This is the case for instance for $T^4$, on which much of our analysis
centers. This is problematic because non-zero integrals of $H_3$ on a
D-brane cycle generate a world-volume tadpole for the dual world-volume
gauge potential $\int_{D7} H_3\wedge A_5$, rendering the configuration
inconsistent \cite{freedwitten} \footnote{ This can be solved as for
the baryonic brane in \cite{wittenbaryon}, by introducing additional
branes ending on the D7-branes \cite{mms}, but this completely
changes the kind of configurations here considered.}. Moreover, in
general, such fluxes along world-volume directions are
quantized, and cannot be diluted away keeping the D7-brane physics
four-dimensional. Hence their presence can lead to qualitatively
large changes in the 4d physics, which may not be well described
with our techniques (which are implicitly perturbative in the flux 
density).

\medskip

For these reasons, we restrict our analysis to situations where fluxes in
the $(2,2')_0$ multiplet are absent. Still, the remaining fluxes
transforming like $(3,1)+(1,3')$ include the most interesting cases, and
lead to interesting and non-trivial effects.

For the computation of soft terms up to the order of interest, it is
enough to consider the leading term in the expansion of $G_3$ in the
normal direction, namely the $z_3$-independent term. Assuming in addition
that the background 3-form flux is independent of the coordinates on the 
4-cycle, as mentioned above, one may obtain the NSNS and RR 2-form 
potentials, in a particular gauge \footnote{We
have chosen a gauge on which all the coordinates are on equal footing.
This is somehow a natural choice for a closed string background,
which is independent of the D-brane configuration under consideration.}
\beqa B_{mn}&=&\frac{g_s}{6i}G^*_{mnp}z^p-\frac{g_s}{12i}
\epsilon_{mnq}S_{\bar{q}\bar{p}}\bar{z}^p+\frac{g_s}{12i}
\epsilon_{mnq}A_{\bar{q}\bar{p}}^*\bar{z}^p-\\ \nonumber & &
-\frac{g_s}{6i}G_{mnp}z^p+\frac{g_s}{12i}
\epsilon_{mnq}S_{\bar{q}\bar{p}}^*\bar{z}^p-
\frac{g_s}{12i}\epsilon_{mnq}A_{\bar{q}\bar{p}}\bar{z}^p\\
\nonumber B_{\bar{m}\bar{n}}&=&-\frac{g_s}{6i}G_{\bar{m}\bar{n}
\bar{p}}\bar{z}^p+\frac{g_s}{12i}\epsilon_{\bar{m}
\bar{n}\bar{q}}S^*_{qp}z^p-\frac{g_s}{12i}
\epsilon_{\bar{m}\bar{n}\bar{q}}A_{qp}z^p+\\ \nonumber &
&+\frac{g_s}{6i}G^*_{\bar{m}\bar{n}\bar{p}}\bar{z}^p-
\frac{g_s}{12i}\epsilon_{\bar{m}\bar{n}\bar{q}}S_{qp}z^p+
\frac{g_s}{12i}\epsilon_{\bar{m}\bar{n}\bar{q}}A^*_{qp}z^p\\
\nonumber B_{m\bar{n}}=-B_{\bar{n}m}&=&\frac{g_s}{12i}
\epsilon_{mpq}S_{\bar{q}\bar{n}}z^p+\frac{g_s}{12i}
\epsilon_{\bar{n}\bar{p}\bar{q}}S^*_{qm}\bar{z}^p-
\frac{g_s}{12i}\epsilon_{\bar{n}\bar{p}\bar{q}}A_{qm}\bar{z}^p-
\frac{g_s}{12i}\epsilon_{mpq}A^*_{\bar{q}\bar{n}}z^p-\\
\nonumber &
&-\frac{g_s}{12i}\epsilon_{\bar{n}\bar{p}\bar{q}}S_{qm}
\bar{z}^p-\frac{g_s}{12i}\epsilon_{mpq}S^*_{\bar{q}\bar{n}}z^p+
\frac{g_s}{12i}\epsilon_{mpq}A_{\bar{q}\bar{n}}z^p+
\frac{g_s}{12i}\epsilon_{\bar{n}\bar{p}\bar{q}}A^*_{qm}\bar{z}^p
\label{integrpuro} \eeqa
The RR 2-forms may be obtained analogously, but we will not need its
explicit expression.

Out of the above components, only those linear in $z^3$, $\bar{z}^3$
will actually be relevant in the computation of soft terms, see appendix 
\ref{calculote}. This suggests that the computation of the 2-form gauge 
potential can be recast in a more compact way
\footnote{An additional advantage is that, as we argue in section
\ref{beyondt4}, this derivation is valid even in situations with 
3-form backgrounds non-constant over the 4-cycle.}, as follows. The 
$(1,3')+(3,1)$ fluxes have always
two legs in $\Sigma_4$. This allows us to associate to each of the above
$SO(4)\times SO(2)$ representations a different 2-form in $\Sigma_4$. In
fact, it is possible to decompose $G_3$ as
\beq G_3=\beta \wedge dz^3 + \beta'\wedge d\bar{z}^3+\gamma
\wedge dz^3 + \gamma' \wedge d\bar{z}^3 \eeq
\label{gsketch}
where $\beta$, $\beta'$ and $\gamma$, $\gamma'$ are
selfdual and anti selfdual two-forms in $\Sigma_4$, namely
\beq *_4\beta=\beta \quad ; \quad *_4\gamma=-\gamma \eeq
(and similarly for the primed forms), corresponding respectively to the
$(1,3')_+$, $(1,3')_-$, $(3,1)_+$ and $(3,1)_-$ pieces of $G_3$.

For clarity we summarize in table \ref{twoforms} the main properties of 
these induced 2-forms.
\begin{table}[htb]
\renewcommand{\arraystretch}{0.70}
\begin{center}
\begin{tabular}{|c|c|c|c|}
\hline Form     & SD/ASD in $\Sigma_4$  & Corresponding $G_3$ rep.  &
ISD/IASD flux \\
\hline \hline
 $\beta$  & SD  &  $(1,3)_+$ & IASD \\
\hline
 $\beta'$  & SD  &  $(1,3)_-$ & ISD \\
\hline
 $\gamma$  & ASD  & $(3,1)_+$  & ISD \\
\hline
 $\gamma'$  & ASD  &  $(3,1)_-$ & IASD \\
\hline
\end{tabular}
\caption{Properties of the 2-forms induced by the flux in $\Sigma_4$.}
\end{center}
\label{twoforms}
\end{table}
The above scalar product between SU(2) triplets will
induce a positive definite product in $\Sigma_4$ given by
\beq \omega_1 \cdot \omega_2=\int_{\Sigma_4}\omega_1\wedge
*_{\Sigma_4}\omega_2\eeq
With all of this, the part of the B-field laying completely in
$\Sigma_4$ is given by
\beq B_2\vert_{\Sigma_4}=-\frac{g_s}{6i}\left((\beta-\beta^*)
z^3+(\beta'-\beta'~^*) z^{\bar{3}} + (\gamma-\gamma^*) z^3 +
(\gamma'-\gamma'~^*) z^{\bar{3}}\right) \label{bkcuatro} \eeq
with $(\beta^*)_{mn}=(\beta'_{\bar{m}\bar{n}})^*$, etc.

In particular, it is important to notice the explicit dependence of these 
components of the B-field on the transverse coordinates. As mentioned, 
these are the only components relevant in the computation of soft terms 
(see appendix \ref{calculote}). 

\medskip

We will further assume that the flux background is purely imaginary
self-dual (ISD) or purely imaginary anti-selfdual (IASD). This assumption
is not compulsory, but simplifies enormously the computations, since in
this situation the equations of motion imply that the dilaton is constant.
It is important to mention that consequently, although our expressions
below include both ISD and IASD components, it is understood that they are
valid just when only ISD or only IASD components are turned on. Namely,
the expressions below do not contain possible interference terms that may
arise when both are present.

\medskip

Concerning the rest of the closed string backgrounds,  we will consider a
general form analogous to that considered above for D3-branes
\beqa
ds^2&=&Z_1^{-1/2}(x)\eta_{\mu\nu}dx^{\mu}dx^{\nu}+Z_2(x)^{1/2}dx^mdx^m
\nonumber \\
\chi_4 & = & \chi(x) dx^0dx^1dx^2dx^3\quad ; \quad
F_5 = d\chi_4+*_{10}d\chi_4
\label{backy1}
\eeqa
Expanding on $z^3,\bar{z}^3$ the warp factor and 5-form flux, we have
\beqa
Z_1^{-1/2}(x)&=&1+\frac{1}{2}K_{mn}(x^a)\  x^mx^n+\ldots \nonumber\\
Z_2^{1/2}(x)&=&1+\frac{1}{2}K_{mn}'(x^a)\ x^mx^n +\ldots \nonumber\\
\chi_4(x)&=&\frac{1}{2}\chi_{mn}(x^a)\ x^mx^n+\ldots
\label{backy2}
\eeqa
with $m,n=8,9$. At this level, $K, K'$ and $\chi$ may depend
on the longitudinal components $z^{1,2}$, $\bar{z}^{1,2}$.
The conditions these backgrounds must satisfy in order to solve Type IIB
supergravity equations of motion are discussed in section \ref{eoms}.

As a last comment, notice that the metric ansatz does not contain any
component mixing the $z_3$ coordinate with the $z_1$, $z_2$. Hence we take
the local geometry to factorize as $\Sigma_4\times C$, namely we work
on local geometries where the normal bundle is trivial. This restricts us
to $T^4\times C$ and K3$\times C$, although clearly the general technique
may be applied to non-trivial normal bundles.

\subsection{The $D=8$ action}
\label{8daction}

Our starting point will be the $D=8$ DBI and CS actions for the D7-brane.
Myers'  action in the Einstein's frame for a D7-brane is given by
\footnote{ In the CS action we only keep pieces  giving possible
contributions to the soft terms.}
\beqa
S=& -\mu_7 \int d^8 \xi \, STr \left[ \, e^{-\phi}\,
\sqrt{-\det(P[E_{\mu\nu}+E_{\mu i}(Q^{-1}-\delta)^{ij}E_{j\nu}]\,
+\, \sigma F_{\mu\nu})\, \det(Q_{ij})} \, \right] \nonumber\\
 & +\, \mu_7g_s\, \int STr \left(\, P\left[\sigma C_6 F_2 + C_8 - C_6 B_2\right]
\right)
\nonumber
\eeqa
where $i,j$ run over the six directions transverse to 4d Minkowski 
space, $P[M]$ denotes the pullback of the 10d background onto the
D7-brane worldvolume, and
\beqa
E_{MN} & = & g_s^{1/2}G_{MN}-B_{MN} \\ \nonumber
Q^m{}_n & = & \delta^m{}_n+i\sigma\,[\phi^m,\phi^p]\,E_{pn}\\ \nonumber
\sigma & = & 2\pi \alpha'
\eeqa
We will label by  $a,b=1,\bar{1},2,\bar{2}$ and  $m,n=3,\bar{3}$
the longitudinal and transverse indices to the D7-brane worldvolume.
Within the class of backgrounds considered, we have $g_{am}=0$, thus
\beq
P[E_{\mu\nu}+E_{\mu i}(Q^{-1}-\delta)^{ij}E_{j\nu}]=
P[g_{\mu\nu}g_s^{1/2}-B_{ab}\delta_{\mu}^a\delta_{\nu}^b+\delta_{\mu}^a
B_{am}(Q^{-1}-\delta)^{mn}B_{nb}\delta_{\nu}^b]
\eeq
where $Q$ is given by
\beq
Q^i~_j=\delta^i~_j+i\sigma[\Phi^i,\Phi^k](G_{kj}g_s^{1/2}-B_{kj})
\eeq
Expanding the above determinants we arrive to the 8d action in
terms of the NS and RR background. The details of the computation
can be found in the appendix \ref{calculote}. The contributions to
the soft terms are given by
\begin{eqnarray}
S_{soft}=\mu_7g_sSTr\,\int \left[
(Z_1^{-1}(\Phi^3,\Phi^{\bar{3}})Z_2(\Phi^3,\Phi^{\bar{3}})
dvol_{4d}\wedge dz^1 \wedge d\bar{z}^1\wedge dz^2 \wedge
d\bar{z}^2-\right. \\
\nonumber \left. -\frac{1}{2}(B_2\vert_{\Sigma_4}-\sigma F_2)\wedge
*_4(B_2\vert_{\Sigma_4}-\sigma F_2) dvol_{4d}+\right. \\
\nonumber \left. +(C_8-C_6\wedge B_2 + \sigma C_6\wedge 
F_2)\vert_{M^4\times \Sigma_4}\right]+\ldots
\label{DBI8}
\end{eqnarray}
with $B_2\vert_{\Sigma_4}$ given in eq. (\ref{bkcuatro}).
Here the dots refer to derivative couplings and other 8-dimensional
contributions which are not explicitly needed in our computations below.

Writing now $C_8$ and $C_6$ in terms of the two forms induced in
$\Sigma_4$ by the flux, and plugging them, together with
(\ref{bkcuatro}), into the above expression we obtain the eight
dimensional soft lagrangian. We refer again to the appendix 
\ref{calculote} for details. The final
result for the $D=8$ bosonic action in terms of $SU(3)$ flux
components, including the kinetic and quartic terms is
\footnote{Notice that the derivative terms mentioned below are required to 
make the expression local Lorentz invariant in 8d. However, the listed 
terms are the only ones explicitly need for our discussion below.}
\beqa {\mathcal{L}}& = &\mu_7\,g_s\,\sigma^2\, {\rm STr}\,
Z_1^{-1}(\Phi^3,\Phi^{\bar{3}})\, Z_2(\Phi^3,\Phi^{\bar{3}})\, \left[ \,
\frac{1}{\sigma^2}+\partial_{\mu}\Phi^3\partial_{\mu} \Phi^{\bar{3}}- 
\right. \\\nonumber 
&- &\frac{g_s}{18}\, \left(\, \frac{1}{4}(S_{12})^2+
\frac{1}{4}(A_{12})^2-\frac{1}{2}G_{\bar{1}\bar{2}\bar{3}}S_{\bar{3}\bar{3}}-
\frac{1}{4}S_{22}S_{11} \, \right) \, \Phi^{\bar{3}}\Phi^{\bar{3}}+h.c.-
 \\ \nonumber
 &-&\frac{g_s}{18}\, \left[\, \vert G_{\bar{1}\bar{2}\bar{3}}
\vert^2+\frac{1}{4}\vert S_{\bar{3}\bar{3}}\vert^2 +\frac{1}{4}
\sum_{i,j=1,2}(\vert S_{ij}\vert^2+\vert A_{ij}\vert^2)\, \right]\,
\Phi^{\bar{3}}\Phi^3-  \\ \nonumber
& +& \sum_{j,k,p=1,2} \frac{g_s}{3}\epsilon_{3jk} \,
\left[ \,(S_{kp})^*+(A_{kp})^* \right]\,
\Phi^3A^{[j}A^{\bar{p}]}+ h.c.+ \\ \nonumber
 & - & \frac{g_s}{6}\epsilon_{ij3}\, S_{\bar{3}\bar{3}}A^iA^j
\Phi^{\bar{3}}+h.c.-\frac{2g_s}{3}\, G_{\bar{1}\bar{2}\bar{3}}\Phi^{\bar{3}}
A^{[\bar{1}}A^{\bar{2}]}+h.c.-  \\ \nonumber
 &- & g_s[A^a,A^{(b}]
[A^{\bar{b})},A^{\bar{a}}]+\partial_{\mu}A^a\partial_{\mu}A^{\bar{a}} -
\frac{1}{2}g_s([\Phi^3,\Phi^{\bar{3}}])^2-\frac{1}{2}g_s
[A^a,\Phi^3][\Phi^{\bar{3}},A^{\bar{a}}]
\left. \,\, \right] \ +\ ... \nonumber
\label{puro}
\eeqa
where $A^a$ are the components of the  8-dimensional gauge boson along the 
directions longitudinal to the D7-brane and $\Phi^{{3}}$ are (adjoint) 
scalar fields describing the transverse degrees of freedom of the 
D7-brane. To avoid long expressions we have not included here either the 
YM action of $D=4$ gauge bosons or couplings depending on derivatives
of the dimensions longitudinal to the D7-brane. Those terms are understood 
to be included in the dots, but, as we commented before, they will not be 
relevant for our discussion below.

The explicit dependence on the antisymmetric backgrounds (in $SU(3)$
notation) and the warp factor is shown. Concerning the latter, notice that 
for pure ISD or pure IASD fluxes, the equations of motion require the warp 
factor to correspond to a black 3-brane solution, with $Z_1=Z_2$, and the 
warp factor dependence drops from the above expression \footnote{This is 
similar to the cancellation of force in systems of a D7-brane in the 
background of D3- or anti-D3-branes, whose supergravity solution is also 
of the black 3-brane form.}. We will consider this situation in what 
follows.

Some of the terms in (\ref{puro}) correspond to the Yang-Mills action for 
the $D=8$ gauge boson and the kinetic terms for the scalars $\Phi^{{3}}$. 
In addition there are terms which depend on the 3-form fluxes. In 
particular note that the transverse scalars  $\Phi^{{3}}$ get mass terms 
for some non-vanishing fluxes. This is an important point since it means 
that, irrespective of the compactification space, in general 3-form  
fluxes stabilize the positions of the D7-branes. This was already 
discussed in section 5 in \cite{ciu}. Note also that in 
addition there are SUSY-breaking scalar-(vector)$^2$ couplings 
proportional to some of the fluxes. Upon further compactification to $D=4$ 
these terms will give rise to  trilinear scalar couplings, as described 
below.

It may be useful for later purposes to display this result in terms of 
the $SU(2)\times SU(2)' \times U(1)$ flux components
\beqa
\mathcal{L}& =& \mu_7\, g_s\, \sigma^2 {\rm STr}\,
\left[\, \frac{1}{\sigma^2}+\partial_{\mu}\Phi^3\partial_{\mu}\Phi^{\bar{3}}- 
\right. \\
\nonumber &-&
\frac{g_s}{36}\, \left[\, {\mathcal{G}}^*\cdot{\mathcal{G}}^*+(G')^*
\cdot (G')^*\right] \,\Phi^3\Phi^3+h.c.-
\frac{g_s}{18}\, \left(\, {\mathcal{G}}\cdot {\mathcal{G}}^*+G'\cdot
(G')^* \right)\, \Phi^{\bar{3}}\Phi^3-  \\ \nonumber &  - &
\frac{g_s\sqrt{2}}{3} \Phi^3 \left[\, \pmatrix{A^{\bar{1}}&
A^{\bar{2}} \cr } ({\mathcal{G}}^*\cdot
{\vec{\sigma}})\pmatrix{ A^1 \cr A^2 \cr } +
\pmatrix{A^{\bar{1}}& A^2 \cr } ({G'}^*\cdot
{\vec{\sigma}})\pmatrix{A^1\cr  A^{\bar{2}} \cr } +h.c.\, \right]
-
 \\ \nonumber
& - & 
g_s\, 
[A^a,A^{(b}][A^{\bar{b})},A^{\bar{a}}]+\partial_{\mu}A^a\partial_{\mu}
A^{\bar{a}} -\frac{1}{2}g_s([\Phi^3,\Phi^{\bar{3}}])^2-\frac{1}{2}g_s
[A^a,\Phi^3][\Phi^{\bar{3}},A^{\bar{a}}]
\left. \,\, \right] \nonumber
\label{puroso4}
\eeqa
where $\vec{\sigma}$ is the vector of Pauli matrices given by
\begin{equation}
{\vec{\sigma}}=\{\sigma_0=\left( \begin{array}{cc}1& 0\\ 0
&-1\end{array}\right), \sigma_x=\left(\begin{array}{cc}0& 1\\ 1&
0\end{array}\right), \sigma_y=\left(\begin{array}{cc}0& -i\\i&
0\end{array}\right) \, \}
\end{equation}

Note the important fact that the effective action depends on the
backgrounds of type ${\mathcal{G}}$ and $G'$, but not on the backgrounds
of type ${\mathcal{G}}'$ and $G$. In terms of the local R-symmetries
$SU(2)\times SU(2)' \times U(1)$ this means that only backgrounds with
negative $U(1)$ charge (i.e. $(3,1)_- +(1,3)_-$) appear in the effective
action. In other words, only fluxes with a particular correlation between
their self-duality properties in the threefold and their duality
properties on the 4-cycle give rise to soft terms, while fluxes with the
opposite correlation lead to cancellation between the DBI and the CS
contributions.

This suggests the following physical interpretation. Consider for
instance fluxes in the $(1,3')_-$ representation. These are ISD in the
threefold, and hence are similar to a D3-brane distribution (rather than
a ${\ov {D3}}$-brane one). On the other hand, they correspond to a
selfdual NSNS 2-form on the D7-brane volume, which induces a ${\ov
{D3}}$-brane charge on the latter \footnote{Notice that our convention 
at this point differs from the usual one.}. The non-cancellation of the 
interaction
between the induced ${\ov {D3}}$-brane on the D7-brane with the
effective D3-brane distribution in the bulk leads to non-trivial soft
terms. A similar argument may be applied for fluxes in the $(3,1)_-$, which
are IASD in the threefold, and anti-selfdual on the 4-cycle. Finally for fluxes
in the $(3,1)_+$ or $(1,3')_+$ the induced charges in the threefold and the
D7-brane volume are of the same kind, leading to cancellations between DBI
and CS contributions.

One should in principle also compute the 8d action for the
world-volume fermion fields. However, it will be simpler to do it directly
in terms of the 4d action, exploiting the symmetries of the system, as we
do below.

\subsection{The $D=4$ action and soft terms}

\subsubsection{Generalities}
\label{gene}

Up to now our discussion has been quite general, and independent of the
4-cycle wrapped by the D7-brane (although as discussed, certain 8d
derivative terms may be necessary to properly describe some situations).
To proceed with a general $\Sigma_4$ we would have to  make a Kaluza-Klein
reduction and compute the massless spectrum, which is very model
dependent.

In this process, even in the absence of fluxes, the non-trivial geometry
determines the number of massless 4d scalars as the number of zero modes
of 8d scalars and internal components of the gauge fields. A prototypical
example is that the number of Wilson line moduli is given by the number of
harmonic 1-forms on $\Sigma_4$ \footnote{If the D7-branes carry a
non-trivial gauge bundle on $\Sigma_4$, the counting of bundle moduli
is more involved. However, we will not consider this situation.}. This is
very often vanishing, but is four for $T^4$. In addition, the 8d complex
scalar fields has a number of zero modes given by the the number of
independent sections of the normal bundle (counted by $h^{2,0}(\Sigma_4)$).
This may be zero in particular cases of rigid 4-cycles (like D7-branes on
the exceptional $P_2$ in the blow-up of the $\IZ_3$ orbifold singularity
${\cal O}_{-3}(\IP_2)$, or on complex delPezzo surfaces). In other 
situations there 
may be multiple massless scalars, associated to different zero modes of 
the 8d scalar. One example is to consider a compactification K3$\times 
T^2$, and the D7-branes wrapped on $\Sigma_4=C_g\times T^2$, with $C_g$ a 
holomorphic genus $g$ Riemann surface in K3. The number of 4d scalars is 
given by the dimension of the moduli space of holomorphic genus $g$ 
Riemann surfaces in K3, given by $g$.

In cases where the normal bundle is trivial, like $T^4\times C$
or K3$\times C$, there is one massless complex scalar $\Phi_3$. It
corresponds to a zero mode of the 8d massless scalars, with
constant profile on $\Sigma_4$. This will be important for our discussion
below.

Any mode which is made massive by the compactification has a typical mass
scale of $1/R$. This is, in the large volume regime of validity of the
description, much larger than the typical flux-induced mass scales, of
order $\alpha'/R^3$. Hence we are interested in describing the effect of
fluxes on the light modes of the configuration, which are the relevant
degrees of freedom from the 4d perspective. There are two sources for these
effects. First, the presence of the fluxes introduces new 8d derivative
couplings of the 8d fields; these would enter into the kinetic energy
operator of the internal space for these fields, and could induce e.g.
non-trivial masses for their former zero modes.
Second, there are 8d non-derivative terms, like 8d mass terms for the
8d $\Phi^3$ scalar, which certainly modify the effective action of the
corresponding 4d zero modes, leading to non-trivial soft terms.
Full determination of these effects  however requires a precise knowledge
of the internal profile of these zero modes, and hence a particular choice
of 4-cycle.

In this section we center on D7-branes on $T^4$. A similar discussion for
D7-branes on K3 is carried out in section \ref{ktres}.

\subsubsection{The $T^4$ case: bosonic sector}

In this section we center on the particularly simple case of
D7-branes wrapped on $T^4$ on $T^4\times C$, where
we moreover assume the background to be independent of the 4-cycle
coordinates (another interesting and tractable case, namely $\Sigma_4=$K3 
is described in section \ref{ktres}). This may be also trivially 
generalized to the orbifold $(T^4\times C)/Z_N$ case, which can lead 
to chiral fermions and semirealistic models, see section \ref{aplica}.

In the toroidal case the 4d massless modes have constant profile in
the internal 4-cycle. The final spectrum in the absence of fluxes
corresponds to an $N=4$ theory. The $D=8$ gauge bosons $A^a$ give rise to
two more complex adjoint scalars $\Phi^{1,2}$, associated to Wilson
line degrees of freedom.

Integrating eq.(\ref{puro}) over the four toroidal dimensions it is easy
to obtain the $D=4$ bosonic action
\beqa
\label{fourdim}
{\mathcal{L}}&=&\Tr 
\left\{ \, \partial_{\mu}\Phi^m\partial_{\mu}\Phi^{\bar{m}}-
\right.\\ \nonumber
&-&\frac{g_s}{18}\, \left(\, \frac{1}{4}[(S_{12})^*]^2+\frac{1}{4}
[(A_{12})^*]^2-\frac{1}{2}(G_{\bar{1}\bar{2}\bar{3}})^*
(S_{\bar{3}\bar{3}})^*-\frac{1}{4}(S_{22})^*(S_{11})^* \,
\right)\, \Phi^3\Phi^3+h.c. \\ \nonumber
& -& \frac{g_s}{18}\, \left(\, \vert G_{\bar{1}\bar{2}\bar{3}}
\vert^2+\frac{1}{4}\vert S_{\bar{3}\bar{3}}\vert^2 +\frac{1}{4}
\sum_{i,j=1,2}(\vert S_{ij}\vert^2+\vert A_{ij}\vert^2)\, \right)
\Phi^{\bar{3}}\Phi^3+  \\ \nonumber
& +& \sum_{k,p=1,2}
\frac{g_s^{1/2}g_{YM}}{6}\epsilon_{ijk}((S_{kp})^*+(A_{kp})^*)
\Phi^i\Phi^j\Phi^{\bar{p}}+h.c.- \\ \nonumber
& - &
\frac{g_s^{1/2}g_{YM}}{6}\epsilon_{ij3}S_{\bar{3}\bar{3}}
\Phi^i\Phi^j\Phi^{\bar{3}}+h.c.-\frac{g_s^{1/2}g_{YM}}{9}
G_{\bar{1}\bar{2}\bar{3}}\epsilon_{\bar{i}\bar{j}\bar{k}}
\Phi^{\bar{i}}\Phi^{\bar{j}}\Phi^{\bar{k}}+h.c.-
\\ \nonumber
& -& g_{YM}^2[\Phi^i,\Phi^{(j}][\Phi^{\bar{j})},\Phi^{\bar{i}}]\,\,
\left. \right\} \nonumber
\eeqa
Here one defines
$g_{YM}^2\ =\ g_s(2\pi)^5(\alpha')^2{\mathcal{V}}^{-1}$,
where ${\mathcal{V}}$ is the volume of the 4-torus

\subsubsection{The $T^4$ case: fermionic sector}

In the same way we one may in principle obtain the soft terms for the fermionic 
fields starting with the supersymmetrized DBI+CS actions in 8d. However 
the computation can be simplified carrying it out in 4d, and exploiting 
the symmetries of the system.

The four-dimensional fermionic action in the $T^4$ case can be
obtained from the dimensional reduction of the supersymmetrized 10d action
\cite{fermions}, up to quadratic order in the fermions, since we are
interested in the fermion masses.  Encoding the 4d fermions in a 10d
Majorana-Weyl spinor $\Theta$ formed by the massless Ramond open string
states, the $SO(3,1)\times SO(4)\times SO(2)$ invariant bilinears
involving the 3-form flux are
\beqa
\label{bilinear}
C\bar{\Theta}\Gamma^{mnp}\Theta[2ab(Re~G_3)_{mnp}&-&
(a^2-b^2)(Im~G_3)_{mnp}]
+\\ \nonumber
&+&C'\bar{\Theta}\Gamma_{(7)}\Gamma^{mnp}\Theta[(a^2-b^2)(Re~
G_3)_{mnp} -2ab(Im~G_3)_{mnp}] \eeqa
where $C$ and $C'$ are two constants, and
\beq
\Gamma_{(7)}=\frac{-i}{8!}\epsilon_{ijklmnpq}
\Gamma^i\Gamma^j\Gamma^k\Gamma^l\Gamma^m\Gamma^n\Gamma^p\Gamma^q
\eeq
with $i\ldots p$ running from 0 to 7. Also, $a$ and $b$ fix the embedding
of the D7-brane supersymmetry in the 10d ${\cal N}=2$ IIB supersymmetry
by means of $\theta_1=a\sigma\Theta$, $\theta_2=b\sigma\Theta$, where
$\theta_1$, $\theta_2$, are the two Majorana-Weyl spacetime spinors of
${\cal N}=2$ supergravity and $a^2+b^2=1$. We work in the choice
$(a,b)=(1,0)$.

The first term in (\ref{bilinear}) arises from contributions from
the DBI piece of the eight dimensional supersymmetric action, whereas the
second piece comes from CS contributions.

The fermionic masses in the four-dimensional action therefore are
given in terms of the four adjoint fermions $\lambda$, $\Psi^i$, $i=1,2,3$,
by
\beqa
{\cal L}_{F}& = &6\sqrt{2}i(C'+C)[(G_{\bar{1}\bar{2}\bar{3}})^*\lambda\lambda+
\frac{1}{2}(S_{\bar{3}\bar{3}})^*\Psi^3\Psi^3+(A_{12})^*\Psi^3\lambda+
\frac{1}{2}\sum_{ij=1,2}S_{ij}\Psi^i\Psi^j]+h.c.+ \nonumber\\
& & +6\sqrt{2}i(C'-C)[G_{123}\lambda\lambda+\frac{1}{2}S_{33}\Psi^3\Psi^3+
A_{\bar{1}\bar{2}}\Psi^3\lambda+\frac{1}{2}\sum_{\bar{i}\bar{j}=\bar{1},\bar{2}}
(S_{\bar{i}\bar{j}})^*\Psi^i\Psi^j]+h.c.\quad 
\label{purofermion}
\eeqa
As a last step, the coefficients $C$ and $C'$ can be determined
easily by supersymmetry arguments. An $S_{{\bar 3}{\bar 3}}$
background is ISD and primitive, and thus preserves unbroken $N=1$
supersymmetry. From (\ref{fourdim}) its scalar superpartner
$\Phi^3$ gets a mass term from such a background (which is one
component in $(1,3')_-$), hence the fermion $\Psi^3$ should get an
equal mass, which fixes $C+C'=-ig_s^{1/2}/72$. One may show from an 
analogous argument applied to $S_{\bar{1}\bar{2}}$, etc. that the second 
term in (\ref{purofermion}) is absent and hence $C=C'$. So finally, the 
four-dimensional fermionic masses are given by
\beq {\mathcal{L}}_F=\frac{g_s^{1/2}}{6\sqrt{2}}
Tr[(G_{\bar{1}\bar{2}\bar{3}})^*\lambda\lambda+
 \frac{1}{2}(S_{\bar{3}\bar{3}})^*\Psi^3\Psi^3+(A_{12})^*\Psi^3\lambda+
\frac{1}{2}\sum_{ij=1,2}S_{ij}\Psi^i\Psi^j]+h.c.
\label{fermionix}
\eeq
Again, note that only the fluxes transforming like $(3,1)_- +(1,3')_-$
appear, with $SU(2)\times SU(2)'\times U(1)$ R-symmetry relating independently
$\lambda$ with $\Psi^3$, and $\Psi^1$ with $\Psi^2$. In fact, the above fermion 
soft masses may be recast in $SO(4)\times SO(2)$ terms as follows 
\beq 
{\mathcal{L}}_F=\frac{g_s^{1/2}}{12}
Tr[ \,\pmatrix{\lambda \, , \, \Psi^3}i\sigma_y (G'^*\cdot
{\vec{\sigma}})\pmatrix{\lambda \cr  \Psi^3 } +
\,\pmatrix{\Psi^1\, \Psi^2}i\sigma_y ({\mathcal
G}^*\cdot {\vec{\sigma}})\pmatrix{\Psi^1 \cr  \Psi^2 }]+h.c.
\label{fermionixso4} 
\eeq 
\subsubsection{Summary}
\label{summary}

Let us summarize the soft terms on the D7 worldvolume fields. In the
notation defined in appendix I in ref \cite{ciu}, one gets
\beqa
m_{1\bar{1}}^2\ & = & m_{2\bar{2}}^2 \ = 0 \ \ ;\ \
B_{ij}\ =\ 0\ \ ,\ i,j\not=3
\nonumber \\
m_{3\bar{3}}^2\ & =&\
\frac{g_s}{18}\left(\vert G_{\bar{1}\bar{2}\bar{3}}
\vert^2+\frac{1}{4}\vert S_{\bar{3}\bar{3}}\vert^2 +\frac{1}{4}
\sum_{i,j=1,2}(\vert S_{ij}\vert^2+\vert A_{ij}\vert^2)\right)
\nonumber
\\
B_{33}\ &=& \ \frac{g_s}{9}\left(\frac{1}{4}(S_{{1}{2}})^{*2}+
\frac{1}{4}(A_{{1}{2}})^{*2}-\frac{1}{2}
(G_{{\bar 1}{\bar 2}{\bar 3}})^*(S_{{\bar 3}{\bar {3}} })^*-
\frac{1}{4}(S_{{2}{2}})^*(S_{{1}{1}})^*\right)
\nonumber
\\
A^{ijk}\ &=& \ -h^{ijk} {{g_s^{1/2}}\over {3\sqrt{2}}}\
(G_{\bar{1}\bar{2}\bar{3}})^*
 \nonumber
\\
C^{ijk}\ &=& -\frac{g_s^{1/2}}{6\sqrt{2}}
\left[\sum_{l=1,2}h^{jkl}(S_{li}+A_{li})-h^{jk3}(S_{\bar{3}\bar{3}})^* \right]
\nonumber\\
M^a\ &=& \  {{g_s^{1/2}}\over {3\sqrt{2}}}\
(G_{\bar{1}\bar{2}\bar{3}})^* \nonumber\\
\mu _{33} \ & =& \  -{{g_s^{1/2}}\over {6\sqrt{2}}} (S_{{\bar 3}{\bar 3}})^*
 \nonumber\\
\mu _{ij} \ & =& \  -{{g_s^{1/2}}\over {6\sqrt{2}}} S_{{ i}{j}}\  , \ i,j=1,2
\nonumber\\
M_{g}^{3a}\ &=&\ {{g_s^{1/2}}\over {6\sqrt{2}}}
 (A_{{ 1}{ 2}})^*
\label{softgen7}
\eeqa
with
\begin{equation}
h_{ijk}=2\epsilon_{ijk}\sqrt{2}g_{YM,77}
\end{equation}
the Yukawa coupling. Again, recall that the above results provide the soft 
terms induced from pure ISD or pure IASD fluxes, since they do not take 
into account possible interference terms.

It is interesting at this point to compare the results obtained
for soft terms in the case of D7-branes to those on D3-branes.
Note the following points:

$\bullet$ Whereas only IASD fluxes $G_{123}$, $S_{ij}$,
$A_{{\bar i}{\bar j}}$ contribute in the case of D3-branes, both ISD and
IASD fluxes appear in the effective action for D7-branes. In particular
note that the ISD flux $G_{\bar{1}\bar{2}\bar{3}}$ does appear in the case
of D7-branes. This is a very important difference. Indeed, it is known
that this kind of ISD background solves the Type IIB supergravity equations 
of motion in global compactifications. Thus, as we will emphasize in more 
detail later (see section \ref{aplica}), one can build orientifold models 
with D7-branes which are solutions of the classical equations of motion 
with zero vacuum energy but still contain explicit SUSY violating soft 
terms. This is not possible with only D3-branes.

$\bullet$ Only some of the components of the flux contribute to the soft 
terms, while for other components cancellation between the DBI and CS 
pieces of the action occur. This is consistent with the physical 
interpretation of soft terms as arising from interaction of   
lower-dimensional D-brane charges induced on the D7-branes with the 
background flux.

$\bullet$ Whereas in the case of D3-branes the fluxes do not
contribute directly to scalar masses (they only do upon application
of the equations of motion), in the case of D7-branes we see that
flux-dependent scalar masses and B-terms directly appear for the
transverse scalars.

$\bullet$ Note also that in the D7-brane case (at least for the
toroidal case which we are discussing now) the scalars $\Phi^{1,2}$
(Wilson line backgrounds) longitudinal to the D7-branes remain massless at
this level. The underlying reason is as follows. The appearance of 8d mass
terms for the internal components of the gauge fields is forbidden by
8d gauge invariance. Then, in the KK reduction on $\Sigma_4$, our
assumption of translational invariance makes it impossible to obtain
non-zero masses in the compactification. This is different for other
4-cycles, for instance K3, where no massless moduli associated to the
D7-branes are present (even in the absence of fluxes). Moreover, we would like
to emphasize that these Wilson line moduli may be absent in orbifold
compactifications, due to the orbifold projection.

\subsection{Solving the equations of motion}
\label{eoms}

In the above results we have not yet fully imposed that the backgrounds
solve the Type IIB supergravity equations of motion (although some
relations, namely $Z_1=Z_2$ and the dilaton being constant, have been 
taken into account). In the present section, we impose the remaining 
conditions, relating the warp factor and 5-form to the flux background, 
and show that those may be solved without affecting these results.

The relevant equations of motion for the dilaton $\tau $, warping $Z$
and 5-form $F_5$ are (see e.g.\cite{ciu})
\beqa
{{i}\over {2}}\nabla^2\tau+{\mathcal{O}} ((\nabla \tau)^{2})&
={{1}\over {24}}G_{mnp}G^{mnp}\\
\nonumber
2\nabla^2Z&={{g_s}\over {12}}G_{\bar{p}\bar{q}\bar{r}}G^*_{pqr}\\
\nonumber
dF_5&={{ig_s}\over {2}}G_3\wedge G_3^*
\eeqa
One can write
\beq
\frac{1}{24}G_{mnp}G^{mnp}=\frac{1}{2}\left(G_{123}
G_{\bar{1}\bar{2}\bar{3}}+\frac{1}{4}S_{lk}S_{\bar{l}\bar{k}}+
\frac{1}{4}A_{lk}A_{\bar{l}\bar{k}}\right) \ ,\
\eeq
Therefore, the first equation is automatically satisfied since we have 
considered $\tau={\rm const.}$ and our flux background is purely ISD or 
purely IASD. Thus e.g. switching on $G_{\bar{1}\bar{2}\bar{3}}$ but not
$G_{123}$ and so on. This justifies a posteriori the type of fluxes that we
describe in the first subsection of this chapter.

The other two equations yield
\beqa
2\nabla^2Z&={{g_s}\over {2}}\left(\vert G_{123}\vert^2+
\vert G_{\bar{1}\bar{2}\bar{3}}\vert^2+\frac{1}{4}\sum_{ij}
(\vert S_{ij}\vert^2+\vert S_{\bar{i}\bar{j}}\vert^2+\vert
A_{ij}\vert^2 +\vert A_{\bar{i}\bar{j}}\vert^2)\right)
\nonumber \\
6\partial_{[\bar{3}}F_{1\bar{1}2\bar{2}3]}&={{ig_s}\over {2}}
\left(\vert G_{123}\vert^2-\vert G_{\bar{1}\bar{2}\bar{3}}
\vert^2+\frac{1}{4}\sum_{ij}(\vert S_{ij}\vert^2-\vert
S_{\bar{i}\bar{j}}\vert^2-\vert A_{ij}\vert^2 +\vert
A_{\bar{i}\bar{j}}\vert^2)\right)
\eeqa
These are easily solved within the ansatz that the warp factor $A$ and
5-form are independent of the coordinates along the 4-cycle, and depend
only on $z_3$, ${\ov z}_3$. This leads to
\beqa
4K_{3\bar{3}}&= {{g_s}\over {2}}\left(\vert G_{123}\vert^2+
\vert G_{\bar{1}\bar{2}\bar{3}}\vert^2+\frac{1}{4}\sum_{ij}
(\vert S_{ij}\vert^2+\vert S_{\bar{i}\bar{j}}\vert^2+
\vert A_{ij}\vert^2 +\vert A_{\bar{i}\bar{j}}\vert^2)\right)
\nonumber\\
-2\chi_{3\bar{3}}&={{g_s}\over {2}}\left(\vert G_{123}
\vert^2-\vert G_{\bar{1}\bar{2}\bar{3}}\vert^2+
\frac{1}{4}\sum_{ij}(\vert S_{ij}\vert^2-\vert
S_{\bar{i}\bar{j}}\vert^2-\vert A_{ij}\vert^2 +\vert
A_{\bar{i}\bar{j}}\vert^2)\right)
\label{soleoms}
\eeqa
In this way all equations of motion are solved. Note that this is
consistent with the type of closed string background (\ref{backy1}),
(\ref{backy2})
discussed at the beginning of this chapter. In particular, in the
present simple solution the 3-form fluxes give rise to a warp factor and
5-form depending only in the dimensions transverse to the D7-branes.
Note on the other hand that the warp factor and 5-form do not appear in
the expressions for the soft terms found above, and hence solving the
equations of motion does not give extra constraints in the form of the 
soft term Lagrangian. This is related to the fact that D7-branes are not
charged under $C_4$, and that the gravitational effect of pure ISD or pure
IASD fluxes (which have $Z_1=Z_2$) on D7-branes cancel.

These are not the most general class of solutions of the
equations of motion, and choosing $Z$ and $F_5$ not depending on
$z^1$, $z^2$ is rather a simplifying assumption in order to
facilitate the computation of the soft terms on the D7-brane worldvolume.
General features beyond this simplified situation are described in
section \ref{beyondt4}.

Note that in the simpler computation of soft terms on D3-branes in
section \ref{dthree} \cite{ciu}, the problem can be solved for arbitrary
coordinate dependence for the warp factor and 5-form (or even the 
dilaton). If one however replaces the above more restricted background, 
constant over the four coordinates, the equations of motion determine the 
masses for the scalars (and not only its trace). In particular from 
eq.(\ref{softgen}) we obtain that the only non-vanishing scalar mass is 
that of the scalar $\Phi_{(33)}^3$ parametrizing the D3-brane position in 
the third complex dimension
\beq
m_{(33)}^2 \ =\
\frac{g_s}{2}\, \left[ \, \vert G_{123} \vert^2\, +\,
\frac{1}{4}\, \sum_{ij}\, (\, \vert S_{ij} \vert^2\, +\,
\vert A_{\bar{i}\bar{j}}\vert^2\, )\, \right]
\label{genmass7}
\eeq
The masslessness of $\Phi^1$, $\Phi^2$ simply means that there is no 
preferred position for the D3-brane in this restricted situation of 
translational invariance along $T^4$

\subsection{$T^4$ with variable warp factor}
\label{variable}

We would like to conclude with some remarks on the results in the slightly 
more general (and more realistic) situation where the background (namely 
the warp factor and 5-form) varies along the 4-cycle on which the 
D7-branes wrap. This may be the case even for $T^4$, as long as the local 
background satisfies the corresponding periodicity conditions. In fact, 
one expects this to be the situation in any compact model, since different 
ingredients, away from the D7-branes (like distant O3-planes, D3-branes or 
other 7-branes) in general distort the warp factor at the location of the 
D7-branes, so that it it not constant over $T^4$.

Clearly, the computation of the 8d action, including the effects of the 
backgrounds, is very similar to the above one, and simply differs in the 
appearance of new derivative couplings, etc. The KK compactification 
proceeds as above, the only difference arising from the fact that the 
backgrounds cannot be pulled out of the integrals over the 4-cycle. This 
implies that the resulting 4d soft terms have exactly the same structure 
as the above computed ones, but their coefficients are integrals of the 
backgrounds over $T^4$, suitably convoluted with the internal wavefunction 
of the 4d zero modes. Since for $T^4$ the latter are constant, the 
coefficients of the soft terms are simply the average values of the 
corresponding background over the internal 4-cycle. Hence, all the above 
expressions remain valid, with the understanding that the constant flux 
coefficients now correspond to their average values over the internal 
4-cycle.

One may think that the introduction of varying backgrounds may lead to 
qualitative changes in the special sector of the Wilson line scalars. 
Indeed our arguments above for their masslessness relied on 8d gauge 
invariance, plus the trivial KK reduction over $T^4$. It could be expected 
that a varying background would introduce new derivative couplings 
for the 8d gauge bosons, leading to non-trivial 4d mass terms for these 
Wilson line degrees of freedom. However this expectation is too naive, 
and not correct: the number of massless 4d scalars from Wilson line 
degrees of freedom is determined topologically, as the first Betti number 
of the internal 4-cycle, i.e. as the number of zero modes of the laplacian 
for 1-forms. The appearance of additional derivative couplings corresponds 
to a change in this internal laplacian, but which cannot change its number 
of zero modes. This implies that the internal 
wavefunction of the 4d Wilson line scalars rearranges in such a way that 
the latter remain massless
\footnote{Note that this implies that the coefficient of the scalar 
trilinear soft terms is not exactly the average value of the flux, but 
rather has a non-trivial convolution with the corrected internal profiles 
for Wilson line scalars.}. Hence, this relevant feature of the toroidal 
case is not changed by considering more general backgrounds.

\section{ Soft terms at $D3-D7$-brane intersections}
\label{dthreeseven}

We will now consider the combined system of D3-branes and D7-branes,
at equal position in the common transverse third complex plane,
$\langle\Phi_{33}^3\rangle=\langle\Phi_{77}^3\rangle$.
The effect of fluxes in the new sector of 37 fields is essentially
different from the previous cases, and notoriously more difficult, since
there is no analog of the DBI+CS action that describes the coupling of
these fields to a general supergravity background.
Nevertheless, the system has a large enough symmetry to constrain the
possible lower-dimension terms, so that with additional partial
information we are able to fully determine them
\footnote{It may be 
possible to exploit the alternative approach in \cite{lmg} to achieve 
these results.}. This is our purpose in
the present section.

Since 37 states are localized at points in the internal manifold, one may
carry out the analysis with a local configuration, without full
information of the 4-cycle spanned by the D7-branes. Nevertheless, for
consistency with our previous discussion we consider flux backgrounds of
the kind there considered.

The local configuration is described by a set of D3-branes, and a set of
D7-branes spanning the coordinates $z_1$, $z_2$, in a (suitably
backreacted) flat space, described by the ansatz (\ref{backy1}). The
local geometric symmetry is $SU(2)\times SU(2)'\times U(1)$, and must be
respected by the soft terms. In the absence of fluxes, the system
preserves 8 supercharges, namely $N=2$ supersymmetry in 4d. The 37 fields
localized at the D3-brane position are four-dimensional, and fill 
out a hypermultiplet of this supersymmetry. Described in terms of $N=1$
supersymmetry
\footnote{For ${\ov{D3}}$/D7-brane systems, to be discussed in section
\ref{antithreeseven}, the system preserves a different $N=1$
supersymmetry. The
discussion is similar, but with an exchange of the roles of the two
$SU(2)$ symmetries.} , we have two $N=1$ chiral multiplets
$(\Phi_{37},\Psi_{37})$ and $(\Phi_{73},\Psi_{73})$, in conjugate
representations of the gauge group. Under the $SU(2)\times SU(2)'\times
U(1)$ symmetry, the complex scalars $\Phi_{37}$ and $\Phi_{73}^*$
transform as $(1,2')_0$, whereas the fermions transform as $(1,1)_{\pm
\frac 12}$.

The system has SUSY superpotential couplings of the form
\beq
W_{37} \ =\ \Phi_{77}^3 \Phi_{73}\Phi_{37}\ -\ \Phi_{33}^3 \Phi_{37}\Phi_{73}
\label{superputi}
\eeq
corresponding to the fact that separating the D3- and D7-brane
locations in the third complex plane gives mass to the hypermultiplet at
the intersection.

The possible soft terms involving the fields at the intersections are
very much restricted by the local geometric symmetries of the system.
Since the scalars belong to doublets of $SU(2)'$, bilinears for these
scalars may only come from 3-form backgrounds transforming like
$(1,3')_-$ or $(1,3')_+$. This means backgrounds of type $(G_{{\bar
1}{\bar 2}{\bar 3}}, A_{12}, S_{{\bar  3}{\bar 3}})$ or else
$(G_{{ 1}{ 2}{ 3}}, A_{{\bar1}{\bar 2}}, S_{{ 3}{ 3}})$. In fact, we will
describe below which explicit couplings of these form are indeed present,
using powerful symmetry arguments. However it is interesting
to motivate their  presence from an independent point of view, as follows.

Although we lack the general action describing the coupling of 37 fields
to a general supergravity background, there is some partial knowledge
about some of these couplings. Namely, the presence of a NSNS 2-form field
along the D7-brane modifies the boundary conditions of open strings ending
on it, and hence induces a change in the mass of scalars from 37 strings.
Such mass terms can be explicitly obtained by computing the open string
spectrum in the presence of this background, using standard world-sheet
techniques. Considering a general B-field background along the D7-brane,
one may use the $SO(4)$ rotation group to bring it to a block diagonal
form, where only the components $B_{1\bar 1}$, $B_{2\bar 2}$ are turned
on. The contribution of this background to masses of 37 states is
described by the mass term in the 4d action of the system
\beqa
\frac{i}{2\pi\alpha'}(\, B_{1\bar 1}+B_{2\bar 2}\, )\, (\Phi_{73} \, \Phi_{73}^* 
+ \Phi_{37}\Phi_{37}^*) \, + \, {\rm h.c.}
\label{dbi37}
\eeqa
Note that there are no terms involving the 37 fermion fields.
As mentioned above, this describes the leading term coupling the 37 fields
to a general NSNS 2-form background, and plays the role of the DBI piece
of the action for 33 and 77 fields.

According to the general philosophy, we may now replace in the above
general term the B-field that corresponds to a specific 3-form flux
background. Namely, at leading order in the coordinates, we have
\beq
B_{mn}\ =\ -\frac{g_s}{6i}(G_{mnp}\ -\ G_{mnp}^*)x^p
\eeq
In particular using (\ref{integrpuro}), the combination
$(B_{1{\bar 1}}+B_{2{\bar 2}})$ is related to the $A_{12}$ 3-form
flux component, since 
\beq 
B_{1{\bar 1}}\ +\ B_{2{\bar 2}}\ =\
-\frac {g_s}{6i}\left( A_{12}\bar{z}^3 \ -\
A_{\bar{1}\bar{2}}z^{3} \ + \ (A_{12})^*z^{3} \ -\ (A_{{\bar
1}{\bar  2}})^*\bar{z}^3 \right) \label{a12} \eeq 
Replacing this in (\ref{dbi37}), and trading $z^3$ for the D7-brane 
world-volume scalar in the third complex plane, we obtain the trilinear 
soft term 
\beq \frac {g_s}{6} [(A_{12})^*-A_{\bar{1}\bar{2}}] \Phi_{77}^3
\Phi_{73}\Phi_{73}^* \ +\ h.c \label{cuanti} \eeq
And as discussed above, there are no fermion mass terms. Notice
that, as announced, the coupling involves a flux component in the
$(1,3')_-$ representation of $SU(2)\times SU(2)'\times U(1)$. Now,
although we computed the above term for a particular flux
component, we may covariantize the expression with respect to
$SO(4)$ to reach the general result of soft terms induced by a
general NSNS 3-form flux.

Note however that this analysis is suggestive but not conclusive, since
we have no information about the coupling of 37 fields to RR fields (say,
the analog of the CS couplings for 33 and 77 sectors). And these are
crucial, since they may (and do, see later) lead to cancellations 
with terms from coupling to NSNS backgrounds. Hence we conclude that 
the above approach is not sufficient to obtain the complete soft terms 
for general NSNS/RR flux backgrounds.

\medskip

Fortunately, the complete answer can be instead found by fully exploiting 
the symmetries of the system, namely supersymmetry of some particular
backgrounds, and the full $SU(2)\times SU(2)'\times U(1)$
geometric symmetry. This can be exploited to obtain the full soft term
lagrangian for 37 fields for a general flux background, as follows.

Consider the D3/D7-brane system in the particular background 3-form flux
$S_{{\bar 3}{\bar 3}}$, which transforms in the $(1,3')_-$  representation.
This is a primitive (2,1)-form flux, hence the combined system preserves
4d $N=1$ supersymmetry. Indeed, in the 77 sector we have seen from
(\ref{softgen7}) that such a background gives rise to a superpotential mass
term for the chiral multiplet containing $\Phi_{77}^3$, $\Psi_{77}^3$,
\beq
W_{\mu }^{(77)} \ =\    -\frac{1}{2}\mu _{(77)}\Phi_{77}^3\Phi_{77}^3\ =\
{{g_s^{1/2}}\over {12\sqrt{2}}} (S_{{\bar 3}{\bar 3}})^*\Phi_{77}^3\Phi_{77}^3
\eeq
On the other hand the $\Phi_{77}^3$ scalar has a superpotential coupling
as in (\ref{superputi}). Thus the relevant piece of the scalar potential
is
\beq
V\ =\ \vert \mu _{(77)}\Phi_{77}^3 \ -\ \Phi_{73}\Phi_{37} \vert ^2,
\eeq
whose crossed term yields a trilinear coupling
\beq
{{g_s^{1/2}}\over {6\sqrt{2}}} (S_{{\bar 3}{\bar 3}})^*\Phi_{77}^3
\Phi_{73}^*\Phi_{37}^* \ +\ h.c.
\eeq
This should be interpreted as the effect of the 3-form flux background
$S_{{\bar 3}{\bar 3}}$ on 37 fields. Notice that it has a structure
reminiscent of the trilinear couplings determined above, but the present
expression, determined purely from supersymmetry considerations,
automatically contains contributions from both the coupling to the NSNS
and RR backgrounds (and hence accounts for possible cancellations between
them).

Now we may use the full $SU(2)\times SU(2)'\times U(1)$ symmetry to
covariantize the above expression, using the fact that $S_{{\bar 3}{\bar
3}}$ is a component of a triplet $(1,3')_-$ of fluxes, which also
includes $G_{{\bar 1}{\bar 2}{\bar 3}}$ and $ A_{12}$. The complete
expression for the soft terms induced by a general flux in the $(1,3')_-$,
is
\beqa {{g_s^{1/2}}\over {6\sqrt{2}}} \big[ (S_{{\bar
3}{\bar 3}})^*\Phi_{77}^3 \Phi_{73}^*\Phi_{37}^* &+ &2(G_{{\bar
1}{\bar 2}{\bar 3}})^*\Phi_{77}^3 \Phi_{73}\Phi_{37} \ +\ \\
\nonumber &+&(A_{12})^*\Phi_{77}^3\Phi_{37}\Phi_{37}^* \ +\
(A_{12})^*\Phi_{77}^3\Phi_{73}\Phi_{73}^* \ +\ h.c. \big]
\label{tri37} 
\eeqa 
Note how indeed the first trilinear scalar coupling advanced in 
(\ref{cuanti}) reappears here from a completely different argumentation, 
while the second is not present (hence due to a cancellation between the 
NSNS and RR flux contributions, automatically taken care of by our 
analysis). The soft terms written in $SO(4)\times SO(2)$ notation read
\beqa {{g_s^{1/2}}\over {6}} (\Phi_{37}^*\quad
\Phi_{73})({G'}^*\cdot {\vec{\sigma}}) \left(
\begin{array}{c}\Phi_{37}\\ \Phi_{73}^*\end{array}\right) \Phi_{77}^3
\label{threesevenso4} \eeqa

\medskip

We may use a similar strategy to compute the soft terms induced from flux
backgrounds in the $(1,3')_+$ representation. Let us start now from a
pure $S_{33}$, which is primitive (1,2) and hence IASD. This flux
does not preserve SUSY, but it was found in \cite{ciu} (see eq.(3.14) in
that reference) that such type of background gives rise to a
supersymmetric mass term for the chiral multiplet containing
the D3-brane fields $\Phi_{33}^3$, $\Psi_{33}^3$, (see also \cite{gp})
\beq
W_{\mu }^{(33)} \ =\    -\frac{1}{2}\mu _{(33)}\Phi_{33}^3\Phi_{33}^3\ =\
{{g_s^{1/2}}\over {4\sqrt{2}}} S_{{ 3}{ 3}}\Phi_{33}^3\Phi_{33}^3
\eeq
On the other hand the $\Phi_{33}^3$ scalar has a superpotential coupling
as in eq.(\ref{superputi}). Thus in the scalar potential there is a term
\beq
V\ =\ \vert \mu _{33}\Phi_{33}^3 \ +\ \Phi_{37}\Phi_{73} \vert ^2
\eeq
whose crossed term yields a trilinear coupling
\beq
-{{g_s^{1/2}}\over {2\sqrt{2}}} S_{{ 3}{ 3}}\Phi_{33}^3
\Phi_{37}^*\Phi_{73}^* \ +\ h.c.
\eeq
Again, we may now use the $SU(2)\times SU(2)'\times U(1)$ symmetry to
covariantize the above expression, and obtain the general soft terms for
37 fields induced by a general flux in the $(1,3')_+$ representation.
The complete trilinear scalar couplings reads
\begin{eqnarray} 
& & {{g_s^{1/2}}\over {2}}(\Phi_{37}\quad
\Phi_{73}^*)({\mathcal{G}'}\cdot {\vec{\sigma}}) \left(
\begin{array}{c}\Phi_{37}^*\\ \Phi_{73}\end{array}\right)
\Phi_{33}^3=\\ \nonumber
& = &-{{g_s^{1/2}}\over {2\sqrt{2}} }\, \big[ S_{{ 3}{
3}}\Phi_{33}^3 \Phi_{37}^*\Phi_{73}^*\ +\ 2G_{{ 1}{  2}{
3}}\Phi_{33}^3 \Phi_{37}\Phi_{73} \ +\ A_{{\bar 1}{\bar
2}}\Phi_{33}^3\Phi_{73}\Phi_{73}^* \ +\ A_{{\bar 1}{\bar
2}}\Phi_{33}^3\Phi_{37}\Phi_{37}^* \ +\ h.c. \big]\, 
\label{tri37IASD} 
\end{eqnarray}
The expressions (\ref{tri37}) and (\ref{tri37IASD}) provide the
complete set of soft terms induced by a general 3-form flux
configuration (since fluxes in the $(3,1)$ cannot couple to 37
scalars, as discussed above). Note also that, as a general
property, trilinear terms involving the D7-brane coordinates
$\Phi_{77}^3$ are sourced by ISD fluxes, whereas those involving
the D3-brane coordinates $\Phi_{33}^3$ are sourced by IASD fluxes.
This is in full agreement with the results from the low-energy 4d
effective supergravity Lagrangian, in section \ref{effective}.

Finally, notice that no scalar or fermion masses are generated. This is
somehow
expected from our previous experience, as follows. Fields in 37 sectors
can be regarded as moduli of an instanton bundle, parametrizing the
possibility of transforming the D3-branes into finite-size instantons on
the D7-brane world-volume. Hence, their dynamics is reminiscent of other
bundle moduli, like Wilson lines, and do not get masses, for reasons
already explained in section \ref{summary}.

\section{Examples of fluxes}
\label{examples}

Let us now discuss some particular examples
of flux choices.

\subsection{ISD fluxes }

\subsubsection{ (0,3) backgrounds}

Let us consider first the case in which only the $G_{{\bar 1}{\bar
2}{\bar 3}}$ background is switched on. This is a $(0,3)$ ISD form which
is of particular relevance since it was shown in \cite{gkp} that such a
background solves the equations of motion with vanishing cosmological
constant \footnote{Note that the IASD backgrounds considered above do also
solve the supergravity equations of motion in the local context, but
cannot be extended to compact spaces keeping vanishing potential energy.}.
From the results in \cite{ciu} reviewed at the beginning of section
\ref{dthree}, no soft terms appear for the D3-brane worldvolume fields. On
the other hand from the results in previous sections, fields from both 77
and 37 sectors acquire soft terms, as follows
\beq
m_{\Phi_{77}^3}^2\ =\
\frac{g_s}{18}\, \vert  G_{{\bar 1}{\bar 2}{\bar 3}}  \vert^2 \ ;\
M^{(77)}\ =  \  {{g_s^{1/2}}\over {3\sqrt{2}}}\
(G_{{\bar 1}{\bar 2}{\bar 3}})^* \ ;\
A^{ijk}~^{(77)}\ =  \ -h^{ijk}\, {{g_s^{1/2}}\over {3\sqrt{2}}}\
(G_{{\bar 1}{\bar 2}{\bar 3}})^*
\label{dildom7}
\eeq
where $A^{ijk}~^{(77)}$ is the coefficient of the trilinear scalar
coupling proportional to $\Phi_{77}^3\Phi_{77}^2\Phi_{77}^1$,
$h_{ijk}=2\sqrt{2}g_{YM}\epsilon_{ijk}$ is the corresponding
Yukawa coupling, and $M^{77}$ are D7-brane gaugino masses.
The rest of the 77 soft terms vanishes. Note the relations
\beq
A^{ijk}~^{(77)}\ =\ - h^{ijk} M^{(77)} \ ;\  m_{\Phi_{77}^3}^2\ =\ |M^{(77)}|^2
\label{dildom2}
\eeq
In addition there is a trilinear scalar coupling involving the 37 scalars
and the D7-brane scalar $\Phi_{77}^3$
\beq
A_{\Phi_{77}^3(73)(37)}\ =\
-{{g_s^{1/2}}\over {3\sqrt{2}}} (G_{{\bar 1}{\bar 2}{\bar 3}})^*
\eeq
Note that these type of soft terms are somewhat  analogous to the
`dilaton-dominated' type of terms which appear for $D3$-fields in an IASD
(3,0) background \cite{ciu}. However an important difference is that
now we have an ISD (0,3) background, which solves the equations of motion
with vanishing cosmological constant, even in compact models. 

\subsubsection{ ISD $S_{{\bar 3}{\bar 3}}$ flux}

This is an ISD primitive $(2,1)$ form and hence we know from general
arguments that they preserve one unbroken SUSY. Thus they may give rise at
most to additional superpotential interactions. From the analysis in
\cite{ciu}, such type of backgrounds give no soft terms for 33 fields,
which are only sensitive to IASD fluxes. On the other hand, for D7-branes,
a $S_{{\bar 3}{\bar 3}}$ flux gives rise to a superpotential term, see
eqs.(\ref{softgen7})
\beq
W_{\mu }^{(77)} \ =\    -\frac{1}{2}\mu _{(77)}\Phi_{77}^3\Phi_{77}^3\ =\
{{g_s^{1/2}}\over {12\sqrt{2}}} (S_{{\bar 3}{\bar 3}})^*\Phi_{77}^3\Phi_{77}^3
\eeq

\medskip

Note that if both $G_{{\bar 1}{\bar 2}{\bar 3}}$ and $S_{{\bar 3}{\bar 
3}}$ fluxes are present there is an additional soft bilinear B-term
(see eq.(\ref{softgen7})) for the scalars $\Phi_{77}^3$
given by 
\beq
B_{33}\ =\ -\frac{g_s}{18} \ (G_{{\bar 1}{\bar 2}{\bar 3}})^*(S_{{\bar 3}{\bar 3}})^*
\ =\ 2M\mu _{(77)}
\label{b33}
\eeq

It is interesting to point out that the full set of bosonic ISD
soft terms may be combined with the SUSY F-term scalar potential to
obtain a perfect square positive definite scalar potential as
follows
\beq
V_{ISD} \ =\ \vert \
-M_{77}^* {\Phi_{77}^3}^* \ -\
\mu _{(77)}\Phi_{77}^3 \ +\ \Phi_{77}^1\Phi_{77}^2 \ +\
\Phi_{73}\Phi_{37}\ \vert ^2
\ =\ \vert \ -M_{77}^* {\Phi_{77}^3}^*\ +\ \partial_{\phi^3} W\ \vert ^2
\label{positive}
\eeq
with $W$ the full superpotential. Indeed it is easy to check that all 
bosonic soft terms described above from ISD backgrounds are reproduced
in this way. The origin of this positive definitive term may be traced 
back to a piece proportional to $|B_2 \ -\ F_2|^2$ in the DBI action in
eq.(\ref{DBI8}). Indeed, for these backgrounds $B_2\simeq G_{{\bar 1}{\bar 
2}{\bar 3}}{\bar z}^3 + (S_{{\bar 3}{\bar 3}})^*z^3$. Upon replacing
$z^3\rightarrow \phi_{77}^3$, and taking into account that upon dimensional 
reduction $F_2$ gives rise to the SUSY F-term (contributing the 
familiar commutator squared to the scalar potential), the above 
structure is obtained. The appearance of this positive definite scalar 
potential has also a clear interpretation in terms of the no-scale 
structure of ISD backgrounds in the effective supergravity action, as we 
show in section \ref{effective}, see eq. (\ref{psititi}).

The above expression of the scalar potential for D7-brane matter fields as
a positive definite quantity is also closely related to the description of the
above configurations in F-theory. Indeed, a configuration of $n$
D7-branes on a 4-cycle $\Sigma_4$ in a threefold $B_3$, can be described
as F-theory on a 4-fold, elliptically fibered over $B_3$, and with an
$I_{n+1}$ Kodaira degeneration of the elliptic fiber over the divisor
$\Sigma_4$. Equivalently, as M-theory on the same fourfold, in the limit
of shrinking the size of the elliptic fiber (and blowing down the
exceptional divisors associated to the singularity from the degeneration).
The D7-brane geometric moduli, i.e. moduli of the 4-cycle  $\Sigma_4$
correspond to complex structure deformations of the F/M-theory fourfold,
describing the geometry of the degenerate fiber locus. Turning on a
general ISD complex 3-form flux $G_3$ on the IIB picture corresponds to
turning on a general real self-dual 4-form flux $G_4$ in F/M-theory, given
by
\beq
G_4\, =\, {{G_3 \, d{\bar w}}\over{\bar{\tau}-\tau}} \, + \, {\rm h.c.}
\eeq
where $dw=dx+\tau dy$ and $\tau$ is the IIB complex coupling.
Hence $(2,1)$ and $(0,3)$ $G_3$ flux components correspond to $(2,2)$ and
$(0,4)$ (plus h.c.) $G_4$ flux components. In the F/M-theory picture, we
have just a 4-form flux in a CY fourfold. This system has been studied in
\cite{beckers} and is described by a superpotential \cite{gvw}
\beq
W \, = \, \int_{CY_4} \, G_4 \wedge \Omega_4
\eeq
where $\Omega_4=\Omega_3 dw$ is the holomorphic 4-form in the fourfold.
This superpotential leads to a positive definite potential for the
scalars, which in particular include the D7-brane geometric moduli. Hence
this confirms the above result for the scalar potential for
D7-brane moduli in ISD 3-form fluxes. The connection has been very 
explicitly shown in \cite{gktt}. It is very satisfactory to recover the 
F/M-theory picture from considerations of the D7-brane action.

\subsection{IASD  backgrounds}

\subsubsection{ $(3,0)$ fluxes}

For this kind of backgrounds the effect of the flux on D3- and D7-fields is
reversed. Indeed, we saw that a background $G_{{ 1}{  2}{ 3}}$ does not give
rise to any soft terms on the D7-brane worldvolume fields. On the other
hand they give rise \cite{ciu} to soft terms for fields in the 33
sector
 \beq
m_{\Phi_{33}^3}^2\ =\ \frac{g_s}{2}\, \vert G_{123} \vert^2 \ ;\
M^{(33)}\ =  \  {{g_s^{1/2}}\over {\sqrt{2}}}\ G_{123} \ ;\
A^{ijk}~^{(33)}\ =  \ -h^{ijk}\, {{g_s^{1/2}}\over {\sqrt{2}}}\ G_{123}
\label{dildom}
\eeq
where $M^{(33)}$ are the D3-brane gaugino masses. In addition there is
a trilinear coupling involving the D3-brane scalar $\Phi_{33}^3$
and the 37 scalars
\beq
A_{\Phi_{33}^3(73)(37)}\ =\ - \
{{g_s^{1/2}}\over {\sqrt{2}}} G_{123}
\eeq
The rest of soft terms vanish. In analogy with the previous case,
there are no scalar masses for the D3-brane fields $\Phi_{33}^1$, 
$\Phi_{33}^2$ due to our simplifying assumptions on the coordinate
dependence of the background. Again, this would change for more general
backgrounds.

The interesting symmetry of soft terms under the exchange $D3
\leftrightarrow D7$, $G_{{\bar 1}{\bar 2}{\bar 3}}\leftrightarrow
G_{{ 1}{  2}{ 3}}$ may be understood from the effective $D=4$ 
supergravity effective Lagrangian approach, as we describe in section 
\ref{effective}.

\subsubsection{$S_{33}$ flux}

We saw in previous section that such a background does not contribute to
the soft term Lagrangian for D7-brane fields. On the other hand, as we
mentioned above, this kind of flux gives rise to a SUSY superpotential
mass term for D3-brane fields
\beq
W_{\mu }^{(33)} \ =\    -\frac{1}{2}\mu _{(33)}\Phi_{33}^3\Phi_{33}^3\ =\
{{g_s^{1/2}}\over {4\sqrt{2}}} S_{{ 3}{ 3}}\Phi_{33}^3\Phi_{33}^3
\eeq
Again this is the only term from such a flux. This is in agreement with
the symmetry under exchanges $D3 \leftrightarrow D7$ and
ISD $\leftrightarrow$ IASD, mentioned above, see section \ref{effective}.

\section{Fluxes and ${\overline{D3}}-D7$ brane intersections}
\label{antithreeseven}

Some interesting Type IIB compactifications make also use of antibranes.
Thus it is interesting to  consider replacing $D3$-branes by
${\overline {D3}}$'s embedded inside D7-branes. Now, as emphasized in
\cite{ciu} the effect  of fluxes on ${\overline {D3}}$'s is obtained from
that with $D3$-branes by replacing $ISD\leftrightarrow IASD$ fluxes.
The effect of the fluxes on the fields at the ${\overline {D3}}-D7$
intersections may be directly derived from those in $D3-D7$ by
symmetry considerations. Indeed, the system ${\overline {D3}}-D7$
preserves a different $N=2$ supersymmetry  compared to that of D3-D7 
system.
The $(\Psi_1,\Psi_2)$ fermions  on $D7$-branes are now inside a $N=2$
vector multiplet whereas those we labeled as $(\Psi_3,\lambda)$ belong to
a hypermultiplet. Thus considering the $SO(4)=SU(2)\times SU(2)'$
symmetries the vector fermions belong to a $(2,1)$ whereas those in
hypermultiplets  belong to a $(1,2')$. The scalars at the ${\overline
{D3}}-D7$ intersections $\Phi_{3{\bar7}}$ and $\Phi_{7{\bar3}}^*$ form a
an $SU(2)$ doublet $(2,1)$ rather than a $(1,2')$. As a consequence only
$(3,1)_-$ and $(3,1)_+$ can couple to these scalars. The trilinear
couplings for $({\bar 3}7)$ scalars are obtained from those  for $(37)$
scalars by replacing the fluxes as follows
\beqa
(1,3')_-:& &(A_{12},S_{{\bar 3}{\bar 3}}, 2G_{{\bar 1}{\bar 2}{\bar 3}})
\ \longrightarrow
\ (3,1)_-: \quad (-S_{{ 1}{ 2}},S_{{ 1}{ 1}},
S_{{ 2}{ 2}})
 \\ \nonumber
(1,3')_+:& &(A_{{\bar 1}{\bar 2}},S_{{ 3}{ 3}}, 2G_{{ 1}{
2}{ 3}})
\ \longrightarrow
\ (3,1)_+: \quad (-S_{{\bar 1}{\bar 2}},S_{{\bar 1}{\bar 1}},
S_{{\bar 2}{\bar 2}})
\label{3*7tril}
\eeqa
We thus get trilinear couplings to IASD fluxes given by
\beqa
{{g_s^{1/2}}\over {6\sqrt{2}}} \big[
(S_{{1}{ 1}})^*\Phi_{77}^3
\Phi_{7{\bar3}}^*\Phi_{{\bar 3}7}^*\ &+&
(S_{{ 2}{ 2}})^*\Phi_{77}^3
\Phi_{7{\bar 3}}\Phi_{{\bar 3}7} \ -\ \\ \nonumber
&-&(S_{{ 1}{ 2}})^*\Phi_{77}^3\Phi_{{\bar 3}7}\Phi_{{\bar 3}7}^* \ -\
(S_{{1}{ 2}})^*\Phi_{77}^3\Phi_{7{\bar 3}}\Phi_{7{\bar 3}}^*
\ +\  h.c.
\big]
\label{tri3*7}
\eeqa
whereas ISD fluxes yield
\beq
-{{g_s^{1/2}}\over {2\sqrt{2}}} \left[
S_{{\bar  1}{\bar  1}}\Phi_{{\bar 3}{\bar 3}}^3
\Phi_{{\bar 3}7}^*\Phi_{7{\bar 3}}^*\ +\
S_{{\bar 2}{\bar 2}}\Phi_{{\bar 3}{\bar 3}}^3
\Phi_{{\bar 3}7}\Phi_{7{\bar 3}} \ -\
S_{{\bar  1}{ \bar 2}}\Phi_{{\bar 3}{\bar 3}}^3
\Phi_{7{\bar 3}}\Phi_{7{\bar 3}}^* \ -\
S_{{\bar  1}{ \bar 2}}\Phi_{{\bar 3}{\bar 3}}^3
\Phi_{{\bar 3}7}\Phi_{{\bar 3}7}^* \ +\ h.c.
\right]
\label{tri3*7IASD}
\eeq
Note that e.g.  in the interesting case of a ISD $(0,3)$ flux
background $G_{{\bar 1}{\bar 2}{\bar 3}}$ no
soft terms appear for fields at the intersections.

\section{Comparison with effective supergravity Lagrangian approach}
\label{effective}

\subsection{The D3/D7-brane system}

Soft terms may also computed in the formalism in \cite{il,kl,bim}, from 
vevs for auxiliary fields for 4d moduli in the 4d effective supergravity 
action. As described in \cite{ciu,ggjl}, 3-form fluxes breaking SUSY 
correspond to non-vanishing auxiliary fields for the dilaton and/or moduli 
fields of $N=1$ preserving CY Type IIB orientifold compactifications.
Hence, we may use the formalism of the 4d effective action to compute soft 
terms on gauge sectors of these compactifications, and compare the results 
with our above local analysis.

Such compactifications always include a massless complex dilaton chiral 
field
\beq
S\ =\ -i\tau \ \quad ;\ \quad \tau \ =\  C\ +\ i/g_s  \ ,
\label{sfield}
\eeq
whose imaginary part is related to the dilaton and  $C$ is
the type IIB
axion. In addition there will be a number of complex structure
scalars $M_a$ and K\"ahler moduli $T_\alpha$.
Consider for simplicity the dependence on a single
overall K\"ahler modulus $T$,
whose real part gives the
overall radius  of the compactification.
 In the large volume
limit the K\"ahler potential has the well known form
\beq
{K\over {M_p^2}} \ = \ -\log(S+S^*)\ -\ 3\log(T+T^*)
\label{kapot}
\eeq
where $M_p^2$  is the 4d Planck mass squared.
We consider compactifications with a non-trivial background
for
the NSNS and RR field strength 3-forms $H_{(3)}$,$F_{(3)}$. In this
situation an $S$-dependent effective superpotential is  given by
\cite{gvw,superp}
\beq
W\ =\  \kappa_{10}^{-2}
\int G_{(3)}\wedge \Omega  \ \ ,\ \ G_{(3)}\ =\
F_{(3)}-iSH_{(3)} \ ,
\label{superp}
\eeq
where $\kappa _{10}^2={1\over 2}(2\pi )^7\alpha '^4$ is the
$D=10$ gravitational constant and
 $\Omega$ the Calabi-Yau holomorphic 3-form.
One can show that standard supergravity formulae then yield for the
auxiliary fields of $T$ and $S$
\beqa
F^S\ &=\ {1 \over
{M_p^2}}\, (S+S^*)^{1/2}\, (T+T^*)^{-3/2}\,
(\kappa_{10}^{-2})\int G_{(3)}{}^*\wedge \Omega
\nonumber \\
F^T\ &=\ -{1 \over {M_p^2}}\,
(S+S^*)^{-1/2}\, (T+T^*)^{-1/2}\, (\kappa_{10}^{-2})\int G_{(3)}\wedge
\Omega
\label{auxiliary}
\eeqa
Thus, e.g., a flux background of $(3,0)$ type
would then correspond to a non-vanishing auxiliary field for  the complex
dilaton $S$, whereas a $(0,3)$ background  would correspond to the
overall K\"ahler field $T$.
The gravitino mass is given by
\beq
m_{3/2}^2 \ =\  { {e^{K/M_p^2} }\over {M_p^4}}\, |W|^2\ = \
{1 \over {M_p^4}}\, (S+S^*)^{-1} \,
(T+T^*)^{-3}\, (\kappa_{10}^{-4})\, |\int G_{(3)}\wedge \Omega\,|^2 \ \
\label{gravit}
\eeq
Note that for a $(0,3)$ background supersymmetry is broken
with a vanishing c.c. (no-scale structure) whereas for
a $(3,0)$ background there is a positive c.c.
given by
\beq
V_0\  = \   { {\kappa_{10}^{-4}|\int G_{(3)}^*\wedge \Omega|^2} \over
{M_p^2  (S+S^*)(T+T^*)^3} } .
\label{vnot}
\eeq
with zero gravitino mass.

Consider now such  Type IIB backgrounds in the presence of
$D7$-  and $D3$-branes.
Although many of the points we will discuss are
more general, let us consider first
 the case of a toroidal
compactification on a factorized $T^6=T^2\times T^2\times T^2$.
Let us  concentrate on the dilaton and the three
$T_i$, $i=1,2,3$ K\"ahler moduli which control the size
of the tori. Consider now a system of D3- and/or D7-branes
(wrapping the first two tori). This is the simplest configuration
that we studied in the first sections of this paper.
Now it was found in ref.\cite{imr} (see also \cite{lrs}) that
the gauge kinetic functions for the $D7$($D3$)
Lagrangian $f_7$($f_3$) as well as the
K\"ahler potential depend on $S,T_i$ as follows:
\beqa
f_3\ &  = & \ S \quad\quad , \quad\quad f_{7}\ =\ T_3 \ ,
\nonumber \\[0.2ex]
K &=& -\log(S+S^* - |\Phi_{77}^{3}|^2)   -  \log(T_3+T_3^* -
|\Phi_{33}^3|^2 )
\nonumber \\[0.2ex]
   & - & \log(T_2+T_2^*-|\Phi_{33}^2|^2 -|\Phi_{77}^{1}|^2 )
-
\log(T_1+T_1^*-|\Phi_{33}^1|^2 -|\Phi_{77}^{2}|^2 )
\nonumber\\[0.2ex]
  &+ & \frac {|\Phi_{37}|^2 + |\Phi_{73}|^2 }
{(T_1+T_1^*)^{1/2}(T_2+T_2^*)^{1/2}  } \ ,
\label{kali}
\eeqa
Focusing only on the dependence on the dilaton $S$ and
overall K\"ahler modulus $T$ this yields for large $T$
\beqa
f_3\ &  = & \ S \quad\quad , \quad\quad f_{7}\ =\ T \ ,
\nonumber \\[0.2ex]
K &=& -\log(S+S^*)   -  3\log(T+T^*)   \\[0.2ex]
&+& \frac { |\Phi_{77}^{3}|^2}{(S+S^*)}
+ \frac {1}{(T+T^*)}\left[ \
(\sum_{a=1}^3 |\Phi_{33}^a|^2)\ +\
(\sum_{b=1}^2 |\Phi_{77}^b|^2)\ +
\ (|\Phi_{37}|^2 + |\Phi_{73}|^2)\ \right] \nonumber
\label{kalifragi}
\eeqa
From this and the results for the auxiliary fields eq.(\ref{auxiliary}) 
one
can compute the SUSY-breaking soft terms for the case of ISD $(0,3)$ and
IASD $(3,0)$ fluxes, which correspond to $(F_S=0, F_T\not= 0)$ and
$(F_S\not=0, F_T= 0)$ respectively by using standard supergravity
formulae. In fact the results may be directly read out from section (7.2)
of reference \cite{imr} \footnote{Specifically the soft terms correspond
to setting $\Theta_i^2=1/3$ in that  reference and $\sin\theta = 0,1$
for the cases  $(0,3)$ and $(3,0)$ respectively.}. Although the results
for the soft terms in terms of the auxiliary field vevs are
given for $D9$ and $D5$ branes, the results equally apply for D3- and
D7-branes  since their low-energy
effective actions   are related by T-duality in the 6 compact
dimensions. The resulting expressions for the D3/D7 system, once we
relate the auxiliary field vevs to the diverse flux components, lead
to the following results

{\bf i) ISD $(0,3)$ background}

This corresponds to $F_S=0, F_T\not=0$.
As expected there are no soft terms for fields from 33 open strings.
For other fields in 37 and 77 sectors, one obtains 
\beqa
m_{\phi_{77}^3}^2\ &=&\ m_{3/2}^2 \ \ ;\ \
m_{\phi_{77}^1}^2\ =\ m_{\phi_{77}^2}^2 \ =\ 0 \\ \nonumber
M^{(77)} \ & =&\  m_{3/2} \\ \nonumber
A_{ijk}^{(77)}\ & =& \ A_{\Phi_{77}^3\Phi_{73}\Phi_{37}}\ =\ -M^{(77)}
\label{soft03}
\eeqa
Using eq.(\ref{gravit}) one can check that this is in agreement
with our results in sections \ref{dseven}, \ref{dthreeseven}.

{\bf ii) IASD $(3,0)$ background}

This corresponds to $F_T=0, F_S\not=0$.
As expected there are no soft terms for fields on the
$D7$ worldvolume. For the $D3$ and $(37)$ fields one gets
\beqa
m_{\phi_{33}^1}^2\ &=&\ m_{\phi_{33}^2}^2\ =\ m_{\phi_{33}^3}^2
\ =\ \frac {V_0}{M_p^2}
 \\ \nonumber
M^{(33)} \ & = &\  \frac {V_0^{1/2}}{M_P} \\ \nonumber
A_{ijk}^{(33)}\ & = &\ A_{\Phi_{33}^3\Phi_{73}\Phi_{37}}\ =\ -M^{(33)}
\label{soft30}
\eeqa
where $V_0$ is given in (\ref{vnot}). The results for gaugino and 
trilinear couplings agree with the results of sections \ref{dseven}, 
\ref{dthreeseven} after substituting $V_0$. However, the results obtained 
for scalar masses are different. As discussed in \cite{ciu}, this 
disagreement is not surprising, since our formulae from the local analysis 
apply to situations with 4d Minkowski space, which in global models are 
only obtained for ISD backgrounds.
Notice also that in the local analysis only $\Phi_{33}^3$ is massive, as a 
consequence of our very restrictive ansatz for the warp factor and 5-form. 
It is clear that in the presence of additional objects, and indeed in any 
complete global compactification, the warp factor and 5-form have a 
non-trivial dependence on all coordinates. This will lead to 
non-trivial mass terms for all D3-branes scalars, as described in the 
general ansatz in \cite{ciu}, in agreement with the result obtained using 
the 4d effective action.

\medskip

We already mentioned in section \ref{examples} that soft terms for 
ISD backgrounds lead to a positive definite scalar potential 
(\ref{positive}) which includes all bosonic soft terms.
This is easy to understand from the effective supergravity
scalar potential. The general form of the later is
\beq
V\ =\ e^{K}\left(  g^{i{\bar j}}(D_i W)( {\bar D}_{\bar j}{\bar W})
\ -\ 3|W|^2\right) \ +\ D-{\rm term}
\label{crem}
\eeq
where the indices run over all the chiral fields and K and W are
the K\"ahler potential and superpotential of the theory. Here
 the K\"ahler covariant derivative is given by
\beq
D_iW\ =\ \partial_iW \ +\ W \partial_i K
\label{cov}
\eeq
If we have $F_S=0, F_T\not=0$ it is easy to see from eq.(7.8)
that the only non-vanishing scalar potential left is that due
to the auxiliary field of the $\phi_{77}^3$ chiral multiplet,
i.e.
\beq
V_{ISD}\ =\ e^{K}\left(  g^{3{\bar 3}}(D_3 W)( {\bar D}_{\bar 3}{\bar W})
\right)
\label{positron}
\eeq
where the index 3 refers to the scalar field $\phi_{77}^3$.
After rescaling the matter fields to canonical value one
gets
\beq
V_{ISD}\ =\
\vert \ -M_{77}^* {\Phi_{77}^3}^*\ +\ \partial_{\phi^3} W\ \vert ^2
\label{psititi}
\eeq
which reproduces the result in (\ref{positive}).

\medskip

We have already noted in different places the apparent symmetry
of SUSY breaking soft terms under the  exchange D3 $\leftrightarrow$ D7, 
$G_{{\bar 1}{\bar 2}{\bar 3}}\leftrightarrow G_{{ 1}{  2}{ 3}}$. This is 
easy to understand from the effective action point of view. Consider the 
gauge kinetic functions and K\"ahler potential in eq.(\ref{kali}) 
corresponding to D7-branes transverse to the third
complex plane and a set of $D3$-branes.
Under a T-duality along the first two complex planes
one has
\beqa
& D7_3 &\ \ \longleftrightarrow \ \ D3 \\ \nonumber
& S & \ \ \longleftrightarrow \ \ T_3    \ .
\label{tduality}
\eeqa
T-duality is thus the origin of the symmetry  of the gauge kinetic functions and
K\"ahler potential in (\ref{kali}) under those replacements. From
this symmetry it is clear that  SUSY-breaking by $F_{T_3}\not= 0$
(resp. $F_S\not=0$) on $D7$-branes leads to the same soft terms as
$F_S\not=0$ (resp. $F_{T_3}\not= 0$) do on $D3$-branes.

\subsection{Intersecting D7-branes and ISD fluxes}
\label{intersection}

Up to now we have just considered  D7-D3 systems in which one stack of 
parallel branes is located transverse to the third complex plane. More 
generally some interesting compactifications may contain different sets of 
D7-branes which wrap different 4-cycles in a CY, and intersect over 
complex curves. In these situations there is a new sector of open strings 
stretched between different D7-brane stacks, and it would be interesting 
to compute the effect of fluxes on their dynamics. Again we lack a DBI+CS 
description of these sectors, and moreover the local geometric symmetry 
(which is reduced to just $SO(2)^3$) is not powerful enough to determine 
the soft terms. However, in this case the effective supergravity 
Lagrangian approach described in this section is still valid, and 
provides a complete well-defined answer.

Let us again consider a simplified situation in which we consider a
factorized toroidal compactification on $T^2_1\times T^2_2\times T^2_3$.
Chiral theories of interest may be later obtained by making
some orbifold projection and/or adding magnetic fluxes on the
worldvolume of the D7-branes (see \cite{lrs} for a recent discussion). 
We will consider three classes of D7-branes, denoted $D7_i$, wrapping 
the 4-tori transverse to the $i^{th}$ complex plane, respectively.
In addition to the matter fields of in $7_i7_i$ , $7_i3$
and $33$ sectors, which we have considered up to now, there are additional
massless hypermultiplets $\Phi_{ij}$  
from the $7_i7_j$ sectors, for $i\not=j$.
Let us consider the case of two different sets of D7-branes $D7_i$, $D7_j$ 
and a set of D3-branes inside their worldvolume.
The 4d superpotential which describes the D3- and D7-branes geometric and 
Wilson line scalars has the form
\beq
W_{D7_iD7_j}\ =\ \phi_{ij}\phi_{ji}\phi_{ii}^k \ -\
       \phi_{ij}\phi_{ji}\phi_{jj}^k \ +\ \phi_{ij}\phi_{37_i}\phi_{37_j}
\label{superpoti}
\eeq
where  $k\not= i,j$ and 
there is no sum on $i,j$ indices. 
Consider now the addition
of a $(0,3)$ ISD background, which corresponds to assuming $F_T\not=0$,
with the rest of the auxiliary fields vanishing. From the
general no-scale potential property described above, only
the auxiliary fields of matter fields on D7-branes contribute to the 
scalar potential. Thus we expect for the scalars at the
$D7_i-D7_j$ intersections to contribute to the scalar potential as
\beq
V_{D7_iD7_j}\ \propto \
\vert \ -M_{77}^* {\Phi_{ij}}^*\ +\ \partial_{\phi_{ij}} W\ \vert ^2
\ =\
\vert -M_{77}^* {\Phi_{ij}}^*\ +\ \phi_{ji}\phi_{ii}^k
\ +\ \phi_{ji}\phi_{jj}^k \ +\ \phi_{37_i}\phi_{37_j} \vert ^2 \ .
\label{soft77}
\eeq
Indeed this may be explicitly verified by using the toroidal
K\"ahler potential for $D7_i$-$D7_j$ system described in section 6
of ref.\cite{imr} and plugging $F_T\not=0$, $F_S=0$. 
One finds the above result with a coefficient of proportionality =1/2
(corresponding to the power 1/2 in the T-dependence of the metric
of the $D7_i$-$D7_j$ fields).
Note that
the scalars $\phi_{ij}$ at the $D7_i$-$D7_j$ intersections become massive
in the presence of a $(0,3)$ background. Thus, as a general conclusion, 
one observes that not only the geometrical moduli
of each D7-brane are fixed by ISD fluxes, but also the open string moduli
associated to the intersections are  fixed. This is a natural 
consequence of these scalars being geometric moduli as well, associated to 
the possibility of recombining the intersecting branes into a single 
smooth one, as we briefly discuss in section \ref{multiple}

\section{ Going beyond the $\Sigma_4=T^4$ case}
\label{beyondt4}

Although in previous sections we have discussed the computation of the
soft terms for D7-branes on $T^4\times C$, the results are slightly more
general. The main simplifying assumptions are given by

i) the metric ansatz, which implicitly assumes that the D7-branes wrap a
4-cycle $\Sigma_4$ with a normal tangent bundle, hence the local
Calabi-Yau geometry is $\Sigma_4\times C$. This on the other hand implies 
that $\Sigma_4$ is Ricci flat, and hence it is $T^4$ or K3.

ii) the fact that the warp factor and 5-form are constant over the
4-cycle. This on the other hand, is not a fundamental restriction, but
rather a technical one which greatly simplifies the computations. As
discussed, we expect a varying background to only slightly modify the
conclusions obtained for a constant one.

iii) specific profiles for the KK reduction. In particular, we have taken
constant profiles for the internal components of the 8d gauge bosons,
leading to the Wilson line moduli $\Phi_1$, $\Phi_2$, and constant
profile for the zero mode of the 8d scalar $\Phi_3$.

In this section we would like to consider how far one may go in relaxing
them, and what results may be obtained at the quantitative or, if not
possible, qualitative level.

\subsection{D7-branes on K3$\times C$}
\label{ktres}

In this section, we argue that the computation can be extended, with
little modification, to the next simplest case, namely D7-branes wrapped
on K3 in the local model K3$\times C$ (see remark i)). Indeed, the local
geometry is consistent with the ansatz for the background 
\beqa
ds^2 & = & Z_1^{-1/2}(x) \eta_{\mu\nu}dx^{\mu}dx^{\nu}
+ Z_2(x)^{1/2} \, g_{mn}^{CY}\,  dx^m dx^m \nonumber \\
& = & Z_1^{-1/2}(x)\, \eta_{\mu\nu}dx^{\mu}dx^{\nu} + Z_2(x)^{1/2}\,
g_{ab}^{K3}\, dx^a dx^b
+ Z_2(x)^{1/2}\, dz_3 d{\ov z}_3 \nonumber \\
\chi_4 & = & \chi(x) dx^0dx^1dx^2dx^3\quad ; \quad
F_5 = d\chi_4+*_{10}d\chi_4
\eeqa
with an expansion (\ref{backy2}) for the supergravity backgrounds.
The background also includes a set of fluxes
\beqa
G_3 & = & \beta \wedge dz_3 \, +\, \beta' \wedge d{\ov z}_3 \, + \,
\gamma \wedge dz_3 \, + \, \gamma'\wedge d{\ov z}_3
\label{k3split}
\eeqa
On K3, the 2-homology is as follows. There is one $(2,0)$- and one
$(0,2)$-form, and 20 (1,1)-forms. The internal product $\omega_1\cdot
\omega_2=\int_{K3} \omega_1 \wedge \omega_2$ endows the 22-dimensional
vector space with signature $(3,19)$, namely there are
3 selfdual and 19 anti-selfdual 2-forms. The above
forms $\beta$, $\beta'$, $\gamma$, $\gamma'$ are linear combinations
of these, according to their duality properties, see table 
\ref{twoforms}, so the expression encodes a large number of components. 
However, most of the physics only depends on the duality properties on the 
4-cycle and on the CY. Hence although $\beta$, $\beta'$, $\gamma$ and 
$\gamma'$ in the above expression contain a larger number of flux 
components, the full computation in appendix \ref{calculote} and in 
sections \ref{dseven}, \ref{dthreeseven} goes through.

Also the equations of motion for the background can be solved consistently 
by considering the warp factor and 5-form to depend only on the directions
$z_3$, ${\ov z}_3$. The result is also irrelevant for the soft term 
lagrangian, hence we skip its expression (which is anyway formally given 
by (\ref{soleoms}), once the latter is translated in terms of $G$, $G'$, 
${\cal G}$, ${\cal G}'$.

The result for the 8d soft term lagrangian is given by expressions 
(3.17) and (\ref{fermionixso4}), with the understanding that 
$G$, $G'$, ${\cal G}$, ${\cal G}'$ correspond to the relevant pieces in 
(\ref{k3split}). Similarly, if there are D3-branes present, the soft terms 
involving the 37 fields are given by (\ref{threesevenso4}), with the same 
understanding.

An important point concerns the KK reduction of this 8d action. Since K3 
has no harmonic 1-forms, the KK reduction of the internal components
of the 8d gauge fields leads to no massless scalars. This can be regarded
as a non-trivial global effect arising from the derivative couplings for
the 8d gauge fields (included in the dots of the expressions for the 8d 
action). The mass scale of 4d massive states in the KK tower
is $1/R$ (where $R$ is a typical length scale of the 4-cycle), which
in the large $R$ regime of validity of the analysis, is much larger than 
the flux-induced masses whose scale is $\alpha/R^3$. Hence, all modes
arising from the internal components of the gauge bosons are not relevant
for the flux physics, and can be ignored in the analysis below. The same
conclusion follows for the supersymmetric partners $\Psi_1$, $\Psi_2$ of
these scalars, which get a large supersymmetric mass due to
compactification effects.

On the other hand, there is one massless scalar from the KK reduction of
the 8d complex scalar $\Phi_3$. Given the large mass of the massive
states in the tower, we may restrict to this zero mode. As easily follows
from the independence of the background in the 4-cycle, the internal
profile of this zero mode is constant over the K3. Hence, despite the
curvature of K3, it is still correct to drop the derivative terms in the
8d action to capture the physics of the massless 4d mode. The final 4d
action for 77 fields can therefore be written \footnote{As discussed in 
section \ref{variable}, since in the K3 case the flux backgrounds are not 
constant, it is understood that the coefficients appearing in the 4d soft 
term lagrangian correspond to the average value of the flux density over 
K3, as follows from the KK reduction.}
\beqa 
{\mathcal{L}}&=&\Tr \big(\partial_{\mu}\Phi^m\partial_{\mu}\Phi^{\bar{m}}- \\ 
&-& \frac{g_s}{36}\, \left(\, {\mathcal{G}}^*\cdot{\mathcal{G}}^*+(G')^*
\cdot (G')^*\right) \,\Phi^3\Phi^3+h.c.-
\frac{g_s}{18}\, \left(\, {\mathcal{G}}\cdot {\mathcal{G}}^*+G'\cdot
(G')^* \right)\, \Phi^{\bar{3}}\Phi^3+  \nonumber\\ \nonumber
& - & \frac{1}{2}g_{YM}^2([\Phi^3,\Phi^{\bar{3}}])^2 + \frac{g_s^{1/2}}{6\sqrt{2}}
Tr[ \,\pmatrix{\lambda \, , \, \Psi^3}i\sigma_y (G'^*\cdot
{\vec{\sigma}})\pmatrix{\lambda \cr  \Psi^3 } + h.c.
\eeqa

This system provides the simplest realization of a D7-brane system where
all D7-brane moduli are lifted. The curvature of K3 eliminates the
possibility of turning on Wilson lines, while the flux induces a mass
term for the moduli associated to transverse motion.

Indeed, this kind of system and the stabilization of all D7-brane moduli
has been discussed in \cite{gktt}, from the perspective of F-theory on
K3$\times K3$. 

\subsection{D7-brane with multiple geometric moduli}
\label{multiple}

As discussed, our analysis does not exactly apply to situations where the
4-cycle has a non-trivial normal bundle. On the other hand, some of the
obtained features can be extrapolated at the qualitative level to these
other situations. Given that the toroidal case is certainly non-generic
concerning the bundle degrees of freedom (e.g. Wilson lines), we
momentarily center on aspects related to the scalars parametrizing motion 
of the D7-brane in the transverse direction.

As discussed in section \ref{gene}, there are local geometries such
that the 4-cycle is rigid, and there are no 4d scalars associated to
D7-brane motion. On the other hand, there are geometries where the 4-cycle
has multiple moduli, leading to several 4d complex scalars, which are
massless in the absence of fluxes. From the viewpoint of the 8d dynamics,
these 4d fields are associated to the zero modes of the laplacian operator
for the 8d scalar field $\Phi^3(x^a)$. We center in this latter case, 
since it is richer from the viewpoint of soft terms.

In the presence of fluxes, one may in principle operate as in section 
\ref{dseven},
allowing for a more general ansatz for the local metric, mixing directions
longitudinal and normal to the 4-cycle. This will induce additional terms
in the 8d action, which we have not considered. On the
other hand, the terms we have computed are present in these situations as
well, so that we again conclude that the 8d scalar $\Phi^3$ acquires an 8d
mass term induced from the fluxes. This term radically modifies its KK
reduction on $\Sigma_4$, in particular eliminating all previously existing
zero modes. Hence we conclude that all geometric moduli for
D7-branes are stabilized by 3-form fluxes. A more quantitative estimate
of these masses can be obtained by splitting the 3-form flux in components
$G$, $G'$, ${\mathcal G}$, ${\mathcal G}'$, according to their
duality properties on the threefold and on $\Sigma_4$, as done in the 
case of K3.

We would like to conclude by mentioning that it is relatively
simple to consider examples of D7-branes with several geometric
moduli. A particularly simple case is to consider a factorized 
six-torus $T^6=T^2\times T^2\times T^2$, and stacks of D7-branes (labeled 
by an index $a$) wrapped on say the third two-torus, times a 2-cycle, 
product of 1-cycles $(n_a^i,m_a^i)$ in $T^2\times T^2$ (where $i=1,2$). 
Each D7-brane has one position modulus (plus Wilson line moduli, which we 
ignore). In addition, two D7-branes intersect a number of times 
$\prod_i(n_a^i m_b^i-m_a^i n_b^i)$. By suitably choosing the geometry, 
the configuration is supersymmetric and at each intersection there is a
massless hypermultiplet. A particular simple example is to consider two 
D7-branes along $(1,0)\times (1,0)$ and $(0,1)\times (0,1)$, which, for 
rectangular two-tori and with some relabeling of coordinates, corresponds 
to the configuration of intersecting D7-branes discussed in section 
\ref{intersection}.

Complex scalars in this hypermultiplet parametrize the possibility of 
recombining the D7-branes into a single smooth one \footnote{This 
possibility may be constrained from D-term conditions. This represents a 
global obstruction to the cycle recombination. So not all hypermultiplets 
at intersections are properly speaking geometric moduli. On the other 
hand, since the effect of fluxes on fields at intersections is local, we 
expect it to be insensitive to global obstructions.}. A convenient local 
model is provided by two D7-branes, spanning a two-torus times the complex 
planes $x=0$ and $y=0$ respectively. The scalars localized at the 
intersection point parametrize the possibility of recombining the 
D7-branes into a D7- on the complex curve
\beqa xy=\epsilon \eeqa
Hence our schematic analysis of D7-branes on 4-cycles with several moduli
applies to the interesting case of intersecting D7-branes. The fact that 
the mentioned moduli get massive in the presence of fluxes is consistent 
with the results we  found in section \ref{effective} from the effective 
lagrangian approach for $7_i7_j$ fields. Related to this, the fact that 
the bosonic soft terms for all 77 fields (including $7_i7_j$ fields) can 
be encoded in a positive definite scalar potential can be understood from 
the F/M-theory perspective, since all 77 fields, again including $7_i7_j$
fields, can be regarded as geometric moduli of the F/M-theory fourfold 
(the latter fields being associated to recombination of singular loci on 
the base).

\section{ Applications }
\label{aplica}

\subsection{Local MSSM-like models from D7/D3-branes at orbifolds and
soft terms}

In this section we present several explicit examples of a local $D7$-$D3$
configurations at orbifolds which lead to a semirealistic models
with the SM gauge group (or in fact some extensions thereof) on the 
worldvolume of $D7$ branes. The geometry is taken to be $(T^4\times C)/Z_N$ 
with D7-branes wrapping the orbifolded $T^4$ and passing through a 
$Z_N$ singularity in the transverse dimensions. We will show that turning 
on ISD fluxes leads to an interesting structure of SUSY-breaking soft 
terms. The interest in having such   kinds  of local configurations
is that, in a second step, we may embed them  in a complete  F-theory 
compactification with 4-form fluxes turned on obtaining in this way  
consistent softly broken $N=1$ compactifications  with  semirelistic 
physics.

Note that in \cite{ciu} somewhat analogous configurations were  
constructed \footnote{The basic ingredients and techniques for model 
building with branes at singularities are discussed in \cite{aiqu} (see 
also \cite{jei}), based 
on the basic results in e.g. \cite{dm}.}
, although with SM physics residing on ${\overline 
{D3}}$-branes and ${\overline {D7}}$-branes. This was required 
in order to obtain non-trivial soft terms for the SM particles
in the presence of an ISD background. On the other hand, realizing
the SM on D7-branes allows to obtain non-trivial SUSY-breaking
soft terms even in ISD backgrounds, as we have discussed in
previous sections. The present case is thus more satisfactory, in the 
sense that the models may be in principle embedded into a complete 
F-theory compactifications leading to  consistent softly broken $N=1$ 
theories once fluxes are added.

Let us first  consider a local geometry $(T^4\times C)/Z_3$ 
with the $Z_3$ twist given by
\beqa
\theta\quad : \quad (z_1,z_2,z_3) \longrightarrow (e^{2\pi i/3} z_1,
e^{2\pi i/3} z_2, e^{-4\pi i/3} z_3)
\eeqa
and let us  locate a stack of nine $D7$-branes at the orbifold
fixed point in the third complex plane.
These D7-branes carry twisted CP factors given by
\beq
\gamma_{\theta,{ 7}}\ =\ \diag\left(\id_3, \alpha\ \id_2, \alpha^2\
\id_2, \id_2\right)
\eeq
with $\alpha = \exp(i2\pi /3)$. Let us take a factorized torus 
$T^4=T^2\times T^2$. The $Z_3$ action leaves $3\times 3=9$ fixed points in 
$T^4$, which we will denote by $(n,m)$, with $n,m=0,\pm 1$. In order to 
reduce the gauge symmetry let us now add one (quantized) Wilson line
with Chan-Paton (CP) matrix
\beq
\gamma_{W,{7}}\ = \ \diag\left( \id_7, \alpha , \alpha^2\right)
\eeq
around (say) the second torus. With the addition of this Wilson line the 
initial gauge group is now broken to $U(3)\times U(2)\times U(2)\times 
U(1)^2$. Note that now the nine fixed points split into three groups of 
three each, feeling different D7-brane CP matrices. In particular the 
points $(n,0)$, $(n,1)$ and $(n,-1)$ feel respectively the CP twists
$\gamma_{\theta,{ 7}}$, $\gamma_{\theta,{ 7}}\gamma_{W,{ 7}}$
and $\gamma_{\theta,{ 7}}\gamma_{W,{ 7}}^{-1}$.

Our configuration should be consistent with local tadpole cancellation 
conditions, which in the present $Z_3$ twist are given by
\beq
Tr\gamma_{\theta,{ 7}}\ +\ 3 Tr \gamma_{\theta,{ 3}}\ =\ 0
\label{tadpole}
\eeq
where $\gamma_{\theta,{ 3}}$ refers to the CP matrices of possible 
D3-branes which may reside at the nine orbifold singularities. The above 
conditions guarantee the cancellation of gauge anomalies in the effective 
field theory. One can easily check that the matrices $\gamma_{\theta,{ 
7}}\gamma_{W,{ 7}}$ and $\gamma_{\theta,{ 7}}\gamma_{W,{ 7}}^{-1}$
are traceless, so that tadpole cancellation conditions is satisfied at the 
six fixed points $(n,1)$ and $(n,-1)$ without the need of adding any
D3-brane on them. On the other hand $\Tr\gamma_{\theta,{7}}=3$ and one 
needs to add D3-branes at the three fixed points $(n,0)$ in order to 
cancel tadpoles. The simplest choice is to add two D3-branes at each of 
the three fixed points, with CP matrices
\beq
\gamma_{\theta,{ 3}_n}\ =\ \diag\left( \alpha, \alpha^2 \right) \ \ ,
n=0,\pm 1
\label{cpd3}
\eeq
which verify $\Tr\gamma_{\theta,{ 3}_n}=-1$. This completes
the description of the local brane configuration.

Although the apparent total gauge group is $U(3)\times U(2)\times 
U(2)\times U(1)^2$ from $D7$-branes and $U(1)^3\times U(1)^3$ from 
$D3$-branes, it turns out that most of the $U(1)$'s become massive due to 
the standard generalized Green-Schwarz mechanism by which pseudoanomalous 
$U(1)$'s combine with RR twisted fields and become massive \cite{iru}.
An anomaly-free combination corresponding to $U(1)_{B-L}$ remains 
massless.Thus the actual gauge group is the simplest left-right
symmetric extension of the gauge group of the SM
\footnote{One can similarly construct 3-generation models
with the SM gauge group. We have chosen here a left-right symmetric
example because the set of  CP matrices is simpler.},
with gauge group  $SU(3)\times SU(2)_L\times SU(2)_R\times 
U(1)_{B-L}$ \footnote{Plus possibly some additional anomaly-free $U(1)$'s, 
which we ignore in what follows.}

From the 77 sector one gets chiral multiplets
transforming with respect to $SU(3)\times SU(2)\times SU(2)\times
U(1)_{B-L}$
like
\beq
Q_L^i\  =\ (3,{ 2},1)_{1/3} \ ;\
Q_R^i\  =\ ({\bar 3},1,2)_{-1/3}\ ;\
H^i\ = \  (1,2,2)_0 \ \ \ ;\ \ i=1,2,3  \ .
\label{esp77}
\eeq
We thus get three generations of quarks together with three sets of
Higgs multiplets. From each of the three  3$_n$7  sectors one gets
chiral multiplets
\beqa
L^n\ & =& \ (1,2,1)_{-1} \ ;\  R^n\ =\ (1,1,2)_{1} \\ \nonumber
D^n\ & = & \ (3,1,1)_{-2/3} \ ;\ {\bar D}^n\  =  \ ({\bar 3},1,1)_{2/3}
\label{esp37}
\eeqa
plus four additional gauge singlets. Finally from each 33 sector one
gets three more singlets. In total the spectrum is that of a Standard 
Model with three generations of quarks, leptons and Higgsses. There are in 
addition three sets of vector-like coloured particles $D^n, {\bar D}^n$,
which may in fact become massive if one of the singlets from the 
33 sector gets a vev. It would be interesting to study the viability of 
this model, but here we 
will be more interested in the structure of SUSY-breaking soft terms which 
may be induced by the presence of fluxes.

Note that the spectrum and perturbative interactions of this type of model 
is simply a $Z_3$ projection of the $N=4$ and $N=2$ actions considered in 
previous section for the case of D7-branes wrapping 4-tori. Thus the
relevant superpotential terms are as follows:
\beqa
W_{77}\ & = & \  \epsilon_{ijk}Q_L^iQ_R^jH^k \\ \nonumber
W_{37}\ & = & \ Q_L^3(\sum_n L^n {\bar D}^n) \ +\
Q_R^3(\sum_n R^n { D}^n)
\label{yuk1}
\eeqa
Now let us assume we add a $(0,3)$ ISD background
$G_{{\bar 1}{\bar 2}{\bar 3}}$. From the results we obtained
in section 5 the following non-vanishing soft terms
are obtained for gaugino masses, scalar masses and
trilinear terms
\beqa
M_3=M_L=M_R=M_{B-L}=M\ &=&  \  {{g_s^{1/2}}\over {3\sqrt{2}}}\
(G_{{\bar 1}{\bar 2}{\bar 3}})^* \nonumber \\
 m_{Q_L^3}^2 \ =\  m_{Q_R^3}^2 \ =\  m_{H^3}^2 \ &=&\ |M|^2 \
\nonumber \\
A_{77}\ = \  A_{37}\ &= &\ - h\ M
\label{softlr}
\eeqa
where here $h$ denotes the superpotential  couplings as in 
eq. (9.8).
Note that all soft terms are determined by a single parameter $M$ and that 
only one generation of quarks gets soft scalar masses. This is not however 
a serious problem, since, once gaugino masses are present, the rest of 
the scalars will get a mass  from one-loop diagrams with gauginos in the 
loop.

In this particular 3-generation model the masses of squarks are not 
universal, which in fully realistic models may lead to phenomenological 
problems with too much flavor changing neutral currents (FCNC). 
Furthermore the $Z_3$ projection does not allow for explicit 
supersymmetric masses (a $\mu$-term) which is one of the ingredients of 
the MSSM. These are properties of this particular model, but not generic 
properties of the flux-induced soft terms. 

To exemplify this fact let us now construct a different local D7/D3-brane
configuration in which both features, universal squark/slepton masses and 
a $\mu$-term, appear. The example is of the Pati-Salam type $SU(4)\times 
SU(2)_L\times SU(2)_R$ which is sufficient for our purposes.

We will be very sketchy in the presentation since we just
want to emphasize only the above mentioned facts.
We start now with
a local geometry
$(T^4\times C)/Z_4$ with the $Z_4$ twist
given by
\beqa
\theta\quad : \quad (z_1,z_2,z_3) \longrightarrow (e^{2\pi i/4} z_1,
e^{2\pi i/4} z_2, e^{ \pi i} z_3)
\eeqa
and  locate a stack of twelve  $D7$-branes at the orbifold
fixed point in the third complex plane.
These $D7$-branes carry twisted CP factors given by
\beq
\gamma_{\theta,{ 7}}\ =\ \diag\left(\id_4, \alpha^3\ \id_2, \alpha\
\id_2, \id_2 , -\id_2 \right)
\eeq
with $\alpha =\exp(i2\pi/4)$.
Let us take again a factorized torus $T^4=T^2\times T^2$. The $Z_4$ action
leaves $2\times 2=4$ fixed points in $T^4$,
which we will denote by $(n,m)$, with $n,m=0, 1$.
We add  now  one  Wilson line
with CP matrix
\beq
\gamma_{W,{7}}\ = \ \diag\left( \id_8, -\id_2 , \id_2\right)
\eeq
around (say) the second torus. The local tadpole
cancellation conditions have now the general
form
$Tr\gamma_{\theta,{ 7}} + 2 Tr \gamma_{\theta,{ 3}} = 0$
for each $\theta$-twisted sector. It is easy to check
that all local tadpoles are canceled if we locate
four D3-branes at each of the two fixed points $(n,0)$ with
$\gamma_{\theta,{ 3}_n}\ =\ \diag\left( \alpha, \alpha^3,
-\id_2 \right)$ for $n=0, 1$.
The D7 gauge group is $U(4)\times U(2)_L\times U(2)_R\times U(2)^2$
and there are chiral multiplets from the 77 sector
as follows
\beqa
F_L^i\ & = & \ (4,{\bar 2},1)\ \ ;\ \ F_R\ = \ ({\bar 4},1,2) \ \ ;\ \ i=1,2
\\ \nonumber
H\ & = & \ (1,{\bar 2}, 2) \ \  ; \ \  {\bar H}\ =\ (1,2,{\bar 2})
\label{psmodel}
\eeqa
Thus there are two standard quark/lepton generations corresponding to the 
first two complex planes. From the third (transverse) complex plane, we
get Higgs doublets able to trigger electroweak symmetry breaking, and with 
Yukawa couplings
$\epsilon_{ij}{\bar H} F_L^iF_R^j$ to quarks and leptons.
We will not display the spectrum from (37) sectors
which just give vector-like multiplets with respect to
the Pati-Salam symmetry.
We now turn on  ISD backgrounds corresponding to
$G_{{\bar 1}{\bar 2}{\bar 3}}$ and $S_{{\bar 3}{\bar 3}}$.
 From the results we obtained
in section 3 and 5 the following non-vanishing soft terms
are obtained
\beqa
M_4 & = & M_L=M_R=M\ =  \  {{g_s^{1/2}}\over {3\sqrt{2}}}\
(G_{{\bar 1}{\bar 2}{\bar 3}})^* \nonumber \\
 m_{H,{\bar H}}^2 \ &=&\ |M|^2 \ \ ;\ \  \mu \ = \
-{{g_s^{1/2}}\over {6\sqrt{2}}} (S_{{\bar 3}{\bar 3}})^*
\nonumber \\
A_{77}\ &=&  - h\ M \ \  ;\  \ B\ =\ M\ \mu
\eeqa
where $\mu $ is a SUSY mass for $H,{\bar H}$. Although this $Z_4$
model has only two generations, the above set of soft terms is
quite simple and predictive and have a number of interesting
properties. All soft terms are determined by the fluxes $G_{{\bar
1}{\bar 2}{\bar 3}}$ and $S_{{\bar 3}{\bar 3}}$ or, alternative by
the two parameters $M,\mu $. The squark/slepton masses are
universal and equal to zero. This poses no phenomenological
problem since they all get large masses at the one-loop level, as
has been abundantly analyzed in the SUSY literature. On the other
hand, due to its universality  FCNC transitions are suppressed. As
we said, one of the interesting aspects is that both a $\mu $- and
a $B$-term, which are important ingredients in the MSSM  are
obtained, with the simple prediction $B=M\mu $.

The above examples  show that ISD fluxes on D7/D3-brane systems not only
are promising in order to stabilize the dilaton and complex
structure moduli,  but also in order to obtain consistent
semirealistic Type IIB backgrounds in which fluxes induce
SUSY-breaking soft terms of possible phenomenological interest. A
comment is in order concerning the scale  of the SUSY-breaking
soft terms. In analogy with the case of D3-branes, they have a typical 
scale of $\alpha'/R^3$, with $R$ a typical length of the 
compactification manifold. In situations with homogeneously distributed 
fluxes, this may be recast as $M_s^2/M_P$. As discussed in \cite{ciu} the 
soft term scale may be as low as the weak scale by taking the string scale 
around the intermediate scale $M_s=10^{11}$ GeV. Another possibility 
would be to exploit the effects of a inhomogeneous warping in the internal 
space to obtain alternative suppression factors.

\subsection{Application to KKLT de Sitter vacua}

As mentioned in the introduction, another application of our results 
is related to the proposal in \cite{kklt}, specifically to the proposed 
mechanism to achieve stabilization of K\"ahler moduli. This should occur 
due to non-perturbative contributions to the superpotential (depending on 
K\"ahler moduli), arising from strong infrared dynamics on D7-brane gauge 
sectors or from euclidean D3-brane instantons. A potential problem in 
generating non-perturbative superpotentials from the formed is that 
7-branes may contain too much massless charged matter. Our results for 
soft terms involving D7-brane matter fields show that the presence of ISD 
fluxes  generally give masses to all the D7-brane geometric moduli. This 
indicates that generically D7-brane vector-like matter is massive, thus 
allowing such non-perturbative superpotentials to appear. One explicit 
example is provided by our K3 example in section \ref{ktres}, where 
in the absence of fluxes the D7-brane gauge theory is pure $N=2$ SYM, 
while after introducing supersymmetric fluxes one is left with pure 
$N=1$ SYM, which develops a gaugino condensate. This 
is also in agreement with recent results derived from the F-theory 
perspective in \cite{gktt}, and from generalized calibrations 
\cite{calgen}.

We would like to mention that, as discussed in \cite{gktt}, an analogous
mechanism may occur concerning the second source of non-perturbative 
superpotentials, namely euclidean D3-brane instantons. An 
euclidean D3-brane instanton (a D3-brane wrapped on a 4-cycle) may 
contribute to the superpotential if if has two fermion zero modes. In the 
standard case, this is achieved if the wrapped 4-cycle, once lifted to 
M/F-theory, leads to a 6-cycle of arithmetic genus equal to one 
\cite{witteninst}. However, this argument ignores possible effects 
of the 
flux on the euclidean D3-brane, which could change the situation as 
follows. Consider a D3-brane instanton wrapped on a 4-cycle leading, in 
the absence of fluxes, to a larger number of fermion zero modes. In the 
presence of fluxes, the instanton action may contain flux-induced terms, 
lifting pairs of such fermions zero modes, and leaving just two in 
suitable situations. Such configurations would then contribute to the 
superpotential.

We would like to point out that, beyond the superficial analogy with the 
flux effects on D7-branes (which also wrap 4-cycles, etc), there is a deep 
connection between the two non-perturbative contributions to the 
superpotential, and the role of flux effects in both situations. Namely, 
instantons on the D7-brane gauge theory can be regarded as euclidean 
D3-branes wrapped on the D7-brane 4-cycle (and dissolved into the 
D7-brane). Hence certain non-perturbative effects on the D7-branes can be 
described in terms of such euclidean D3-branes. For instance, the gaugino 
condensate of a D7-brane $N=1$ $SU(n)$ gauge theory can be understood in 
terms of fractional instantons, described by euclidean D3-branes 
whose 6-cycles in the F/M-theory picture are given by exceptional 
divisors of the elliptic fibration degeneration (i.e. the base 4-cycle 
times a collapsed 2-sphere in the elliptic fibration degeneration). Thus, 
the computation of the lifting of D7-brane massless matter, leaving an 
infrared gauge theory with strong dynamics, contains the implicit 
computation of the lifting of euclidean D3-brane instanton fermion zero 
modes from flux-induced effects.

It would be very nice to carry out the computation of the flux-induced 
effects on euclidean D3-branes more explicitly. We hope the techniques 
reported in this paper are useful in this direction.

We would like to end by making  a few comments about the extent to 
which our results for soft terms apply to the scenario in \cite{kklt} to 
construct deSitter vacua in string theory. In the original proposal there 
are essentially three ingredients:

\begin{itemize}

\item One first compactifies Type IIB theory on a CY orientifold
and adds $(0,3)$ ISD fluxes. The superpotential induced by the
flux fixes the dilaton and all complex structure moduli. The
theory has a no-scale structure with a vanishing cosmological constant.

\item In a second step a modulus dependent superpotential
$W(T)$ is introduced. This breaks the no-scale structure
and gives rise to an isolated supersymmetric anti de-Sitter minimum, with
negative cosmological constant. The modulus $T$ is thus fixed.

\item Finally the cosmological constant is increased to a (possibly tiny) 
de Sitter value, by adding a set of anti-D3-branes in the bulk (for other 
possible sources of positive tension, see e.g. \cite{positive}).

\end{itemize}

The kind of soft terms we have computed apply
to the theory obtained after the first step only.
In principle,
using the effective supergravity Lagrangian approach
and including a  non-perturbative superpotential $W(T)$
one can also  compute soft terms in the KKLT vacua.
The results however will depend on the form of $W(T)$   
as well as on the K\"ahler potential and gauge kinetic function
of the effective theory. Furthermore, there will be additional
dependence on the specific way by which we increase
the vacuum energy up to de Sitter space. Preliminary work seems to show 
that, depending on how all these details work out, our specific results 
for soft terms will apply or not to de-Sitter vacua with the ingredients 
in \cite{kklt}. We postpone a systematic discussion of these issues for 
future work.

\vspace*{0.5cm}

\section{Final comments}

Flux compactifications have been introduced as a canonical mechanism to 
lead to stabilization a large number of moduli in string 
compactifications. Interestingly, besides addressing this long-standing 
problem in string theory, they provide an extremely tractable mechanism to 
break supersymmetry in string theory, in a computable regime. In fact, 
despite the seemingly complicated structure of the models, it is possible 
to carry out detailed computations of the flux-induced soft SUSY-breaking 
terms in gauge sectors localized on the volume of D-branes in the models.

In this paper we have described the effects of field strength 3-form 
fluxes on configurations of D3/D7-branes, with D7-branes wrapped on 
4-cycles, in Calabi-Yau compactifications.
The computation and results turn out to be richer and more interesting 
than the similar effects on D3-branes, discussed in \cite{ciu,ggjl}.

We have carried out a complete analysis of flux-induced soft terms on 
D7-branes in local Calabi-Yau geometries, by expanding the DBI+CS 
world-volume action coupled to the flux supergravity background.
For the more complicated sector of open strings stretched between D3 (or 
anti-D3) branes and D7-branes, where a similar general action is not 
known, we have succeeded in exploiting the symmetries of the system to 
fully determine the soft terms.

We have emphasized that in D3/D7-brane systems, a prominent role in 
determining the 4d physics is played by the 4-cycle on which the 
D7-branes wrap. Although our computations have been explicitly carried out 
only in the case of D7-branes wrapped on $T^4$, the correct identification 
of the underlying physical interpretation of the soft terms has allowed 
us to extrapolate the results to other 
situations, like D7-branes on K3, or on 4-cycles with multiple geometric 
moduli.

The results markedly differ from the situation with just D3-branes. 
Indeed, we have found that gauge sectors on D7-branes acquire non-trivial 
soft terms even in the presence of ISD 3-form fluxes. This is interesting, 
since such fluxes lead to consistent string compactifications to 
4d Minkowski space without any runaway potential for K\"ahler moduli. 
Hence the models constitute 4d classical string vacua with zero 
cosmological constant and non-trivial soft terms for the gauge sector on 
D7-branes. 

This implies no contradiction whatsoever with the no-scale 
structure of the 4d effective action. Indeed, we have exactly reproduced 
the soft terms computed in the local analysis from the 4d effective 
action, where the flux background components are identified with auxiliary 
fields of the different moduli chiral multiplets.

This is an 
improvement over the situation with just D3-branes, where, in order to 
obtain non-trivial soft terms,  one must either consider D3-branes in IASD 
fluxes (which in compact models lead to runaway potentials for K\"ahler 
moduli), or anti-D3-branes in ISD fluxes (where now the backreaction of 
the antibranes would induce the potential).

The set of four-dimensional gauge sectors which can be constructed using 
D3/D7-brane configurations is extremely rich, and includes semirealistic 
models, which are nevertheless very tractable (being orbifold projections 
of toroidal and flat space models). Therefore, we believe that D3/D7-brane 
configurations embedded in 3-form flux backgrounds provide a well-defined 
and very interesting starting point for phenomenological studies 
of SUSY-breaking in string theory. Further discussions in this direction 
have been initiated in \cite{fluxedmssm}.

Finally we have also emphasized that our results have an important 
application to the recent proposal in \cite{kklt} to stabilize all moduli. 
In particular, flux effects lifting charged vector-like matter on
D7-brane world-volume help in generating non-perturbative superpotentials for 
K\"ahler moduli. 

Clearly, many questions remain open. For instance, it would be interesting 
to evaluate the impact on soft terms of additional ingredients possibly
present in more involved compactifications (like non-perturbative 
superpotentials for K\"ahler moduli, or additional sources of tension). 
It is very tempting to speculate that, since D-branes are only 
sensitive to the local background around them, and our analysis has 
considered a fairly general such local background (with 4d 
Poincare invariance being the only strong assumption), our results will be 
valid for any model with small enough cosmological constant (much smaller 
than the flux scale). 

Also it would be interesting to extend our analysis to more involved 
configurations of D7-branes, for instance D7-branes carrying non-trivial 
gauge bundles on their world-volume. In the toroidal setup, 
such models are dual to configurations of intersecting D6-branes, which 
have been extensively studied for phenomenological model building. Hence, 
a detailed analysis of the resulting soft terms for these configurations 
would very much improve the possible phenomenological applications of 
fluxes. 

We expect much progress in these and other directions, in the interesting 
interplay of D-branes and fluxes, in particular concerning supersymmetry 
breaking and soft terms.

\medskip

\begin{center}
{\bf \Large Acknowledgments}
\end{center}

We thank J. F. G. Cascales, D. Cremades, S. Kachru, F. Marchesano, F. 
Quevedo and S. Trivedi for useful  discussions. We thank the authors of 
ref.\cite{gktt} and \cite{msw} for informing us of their results 
prior to publication.
L.E.I. and P.G.C. thank CERN's Theory Division for hospitality.
A.M.U. thanks M.Gonz\'{a}lez for patience and support.
This work has been partially supported by the European Commission under
the  RTN contract HPRN-CT-2000-00148 and the CICYT (Spain).
The work of P.G.C. is supported by  the Ministerio de Educaci\'on, Cultura
y Deporte (Spain) through a FPU grant.

\newpage

\appendix

\section{Computation of the D=8 action and dimensional reduction in a $T^4$.}
\label{calculote}

In this section we present the detailed computation of the
expressions (\ref{puro}) and (\ref{fourdim}). The conventions that
we will adopt are exactly the same than in \cite{ciu}, summarized
in the appendix therein.

As we mentioned in the main text, our starting point is Myers'
action for a D7-brane written in the Einstein's frame
\beqa
S=& -\mu_7 \int d^8 \xi \, STr \left[ \, e^{-\phi}\,
\sqrt{-\det(P[E_{\mu\nu}+E_{\mu i}(Q^{-1}-\delta)^{ij}E_{j\nu}]\,
+\, \sigma F_{\mu\nu})\, \det(Q_{ij})} \, \right] \nonumber\\
 & +\, \mu_7g_s\, \int STr \left(\, P\left[\sigma C_6 F_2 + C_8 - C_6 B_2\right]
\right)
\nonumber
\eeqa
where in the CS action we have kept only pieces giving contributions
to the soft terms. For our particular background 
$P[E_{\mu\nu}+E_{\mu i}(Q^{-1}-\delta)^{ij}E_{j\nu}]$ and
$Q^i_{\, j}$ are given by
\beqa P[E_{\mu\nu}+E_{\mu i}(Q^{-1}-\delta)^{ij}E_{j\nu}]& = &
P[g_{\mu\nu}g_s^{1/2}-B_{ab}\delta_{\mu}^a\delta_{\nu}^b+\delta_{\mu}^a
B_{am}(Q^{-1}-\delta)^{mn}B_{nb}\delta_{\nu}^b]\nonumber \\
Q^i~_j& =
&\delta^i~_j+i\sigma[\Phi^i,\Phi^k](G_{kj}g_s^{1/2}-B_{kj})
\nonumber \eeqa
We start by computing the first determinant in the DBI piece of
the action. Neglecting derivative couplings, it can be factorized
between the Minkowski and the 4-cycle pieces as
\beqa
& & det(P[E_{\mu\nu}+E_{\mu i}(Q^{-1}-\delta)^{ij}E_{j\nu}]+F_{\mu\nu})= \nonumber \\
& = & det(g_{\mu\nu}g_s^{1/2}+2g_s^{1/2}\sigma^2D_{\mu}\Phi^3D_{\nu}\Phi^{\bar{3}}
+\sigma F_{\mu\nu})\cdot det(g_s^{1/2}+\sigma F_{ab}-B_{ab}+\\ \nonumber & & +2g_s^{1/2}\sigma^2
D_a\Phi^3D_{b}\Phi^{\bar{3}} + 
B_{am}(Q^{-1}-\delta)^{mn}B_{nb}-\sigma B_{3(a}D_{b)}\Phi^3-\sigma
B_{\bar{3}(a}D_{b)}\Phi^{\bar{3}}) \nonumber \eeqa
with $a,b$ running over the internal coordinates of $\Sigma_4$ and
\beq
D_a\Phi^m=\partial_a\Phi^m+i[A_a,\Phi^m]=i[A_a,\Phi^m]
\eeq
so there can still be possible contributions from the covariant
derivative in the non-abelian case.

Ignoring derivative couplings is justified because the relevant fields in 
the two cases we center on, namely $T^4\times C$ and K3$\times C$, 
have constant profiles in the 4-cycle, as explained in the main text. 
It is however straightforward (but lengthy) to take them into account to 
obtain the complete 8d action, hence we do not provide the result.

We can expand now these two determinants making use of the formula
\beq
det(1+M)=1+Tr~M-\frac{1}{2}Tr~M^2+\frac{1}{2}(Tr~M)^2+\ldots
\eeq
where the dots in this and in further expressions refer to contributions
giving rise to derivative couplings or couplings with dimension higher 
than four in the low energy effective action.

Therefore,
\beqa & & det(P[E_{\mu\nu}+E_{\mu
i}(Q^{-1}-\delta)^{ij}E_{j\nu}]+\sigma F_{\mu\nu})=\\ \nonumber &
= &
-g_s^4[Z_1^{-1}(\Phi^3,\Phi^{\bar{3}})Z_2(\Phi^3,\Phi^{\bar{3}})]^2+
2g_s^4\sigma^2\partial_{\mu}\Phi^3\partial_{\mu}\Phi^{\bar{3}}-
g_s^3\frac{1}{2}(B_{ab}-\sigma F_{ab})(B_{\bar{a}\bar{b}}-\sigma
F_{\bar{a}\bar{b}}) -
\\ & - & g_s^4\sigma^2 F_{\mu a}F^{\mu a}+\sigma^2 g_s^{9/2}[A_a,\Phi^3][A^a,\Phi^{\bar{3}}]
+ig_s^4\sigma(B_{3a}[A^a,\Phi^3]+B_{\bar{3}a}[A^a,\Phi^{\bar{3}}])+\ldots
\nonumber \eeqa
The last term of this expression will survive to the KK reduction
only in cases on which the Wilson line moduli and $\Phi^3$ have
profiles with different parity. This is not the case in
compactifications on 4-cycles trivially fibered in the normal
direction. Therefore, in what follows we will include as well
these terms in the final dots.

The second determinant of the DBI piece is much simpler to compute
\beq det(Q^i_j)=1+\sigma^2([\Phi^3,\Phi^{\bar{3}}])^2g_s \eeq
Putting everything together and Taylor expanding the square root,
we have the following eight dimensional action
\beqa {\cal L}& = &\mu_7g_sSTr\,
Z_1^{-1}(\Phi^3,\Phi^{\bar{3}})Z_2(\Phi^3,\Phi^{\bar{3}})
\left(1+\sigma^2\partial_{\mu}\Phi^3\partial_{\mu}\Phi^{\bar{3}}-\right.
\\ \nonumber & & \left.
-\frac{1}{2}(B_2\vert_{\Sigma_4}-\sigma F_2)\wedge
*_4(B_2\vert_{\Sigma_4}-\sigma F_2)+ \frac{\sigma^2}{2}F_{\mu
a}F^{\mu a} -\right. \\ \nonumber & & \left.
-\frac{1}{2}g_s\sigma^2([\Phi^3,\Phi^{\bar{3}}])^2-
\frac{1}{2}g_s\sigma^2[A^a,\Phi^3][\Phi^{\bar{3}},A^{\bar{a}}] +
(\sigma C_6 \wedge F_2+C_8-C_6\wedge B_2)\vert_{\Sigma_4}+\ldots
\right) \label{casifin} \eeqa
It only remains to plug our background in this expression. From the 
equations of motion of Type IIB supergravity (see e.g. \cite{ps})
one realizes that $C_6$ and $C_8$ (in our conventions) are given 
by
\beqa dC_6 & = & H_3\wedge \left(C_4+\frac{1}{2}B_2\wedge C_2
\right)-*Re G_3\\ \nonumber dC_8 &=& H_3\wedge C_6 - *Re ~ d\tau
\eeqa
Decomposing $G_3$ as in (\ref{gsketch}) and making use of these
expressions, we can integrate the relevant RR and NS field
strengths for our particular background \footnote{We have chosen a
gauge on which all the coordinates transverse to Minkowski are
treated in the same way.} to obtain
\beqa C_6\vert_{M_4\times \Sigma_4}& =
&-\frac{1}{6i}((\beta-\beta^*) z^3 -
(\beta'-{\beta'}^*) \bar{z}^3 - (\gamma-\gamma^*) z^3 + 
(\gamma'-{\gamma'}^*) \bar{z}^3)\wedge d~Vol_{4d}+\ldots \nonumber \\
\nonumber C_8\vert_{M_4 \times \Sigma_4}& = &-\frac{g_s}{36}
\left((\beta z^3 + \gamma'
\bar{z}^3-{\beta'}^*\bar{z}^3-\gamma^*z^3)\wedge(\beta z^3 +
\gamma' \bar{z}^3-{\beta'}^*\bar{z}^3-\gamma^*z^3)-\right. \\
\nonumber & & \left. (\beta' \bar{z}^3+\gamma
z^3-\beta^*z^3-{\gamma'}^*\bar{z}^3)\wedge(\beta' \bar{z}^3+\gamma 
z^3-\beta^*z^3-{\gamma'}^*\bar{z}^3)\right)\wedge d~Vol_{4d}\\
B_2\vert_{\Sigma_4}& =
&-\frac{g_s}{6i}\left((\beta-\beta^*) z^3+(\beta'-{\beta'}^*)
z^{\bar{3}} +
 (\gamma-\gamma^*) z^3 + (\gamma'-{\gamma'}^*) z^{\bar{3}}\right)
\eeqa
with $(\beta^*)_{mn}=(\beta'_{\bar{m}\bar{n}})^*$, etc.

Plugging these expressions in (\ref{casifin}) we finally get
\beqa {\cal L}&=&\mu_7g_s\sigma^2STr\,
Z_1^{-1}(\Phi^3,\Phi^{\bar{3}})Z_2(\Phi^3,\Phi^{\bar{3}})
\left(\frac{1}{\sigma^2}+\partial_{\mu}\Phi^3\partial_{\mu}\Phi^{\bar{3}}+
\right.\\ \nonumber 
& & \left. -\frac{g_s}{36}(2\gamma^*\wedge *_4\gamma'
\Phi^{\bar{3}}\Phi^3+ 2\beta^*\wedge *_4\beta'
\Phi^{\bar{3}}\Phi^3-\gamma'\wedge
*_4\gamma'\Phi^{\bar{3}}\Phi^{\bar{3}} -\beta'\wedge
*_4\beta'\Phi^{\bar{3}}\Phi^{\bar{3}}+h.c.)\vert_{\Sigma_4}\right.
\\ \nonumber & & \left. - \frac{g_s}{3}(
\beta'_{ab}\Phi^{\bar{3}}A^aA^b+
\gamma'_{ab}\Phi^{\bar{3}}A^aA^b+h.c.)-\right. \\ \nonumber & &
\left.
-\frac{g_s^{-1}}{4}F_{ab}F_{\bar{a}\bar{b}}+\frac{1}{2}g_s^{-1}F_{\mu
a}F_{\mu \bar{a}} -\frac{1}{2}g_s([\Phi^3,\Phi^{\bar{3}}])^2-
\frac{1}{2}g_s[A^a,\Phi^3][\Phi^{\bar{3}},A^{\bar{a}}]+\ldots
\right) \eeqa
which in terms of the SU(3) irreducible representations of Table 1
can be rewritten as
\beqa {\mathcal{L}}& = &\mu_7g_s\sigma^2STr
Z_1^{-1}(\Phi^3,\Phi^{\bar{3}})Z_2
(\Phi^3,\Phi^{\bar{3}})\big(\frac{1}{\sigma^2}+
\partial_{\mu}\Phi^3\partial_{\mu}
\Phi^{\bar{3}}-
\\ \nonumber
&- &\frac{g_s}{18}\big(\frac{1}{4}(S_{12})^2+
\frac{1}{4}(A_{12})^2-\frac{1}{2}G_{\bar{1}\bar{2}\bar{3}}S_{\bar{3}\bar{3}}-
\frac{1}{4}S_{22}S_{11}\big)\Phi^{\bar{3}}\Phi^{\bar{3}}+h.c.-
 \\ \nonumber
 &-&\frac{g_s}{18}\big(\vert G_{\bar{1}\bar{2}\bar{3}}
\vert^2+\frac{1}{4}\vert S_{\bar{3}\bar{3}}\vert^2 +\frac{1}{4}
\sum_{i,j=1,2}(\vert S_{ij}\vert^2+\vert A_{ij}\vert^2)\big)
\Phi^{\bar{3}}\Phi^3+  \\ \nonumber
& +& \sum_{j,k,p=1,2}
\frac{g_s}{3}\epsilon_{3jk}((S_{kp})^*+(A_{kp})^*)\Phi^3A^{[j}A^{\bar{p}]}+
h.c.- \\ \nonumber
 & - & \frac{g_s}{6}\epsilon_{ij3}S_{\bar{3}\bar{3}}A^iA^j
\Phi^{\bar{3}}+h.c.-\frac{2g_s}{3}G_{\bar{1}\bar{2}\bar{3}}\Phi^{\bar{3}}
A^{[\bar{1}}A^{\bar{2}]}+h.c.-  \\ \nonumber
 &- & g_s[A^a,A^{(b}]
[A^{\bar{b})},A^{\bar{a}}]+\partial_{\mu}A^a\partial_{\mu}A^{\bar{a}} -
\frac{1}{2}g_s([\Phi^3,\Phi^{\bar{3}}])^2-\frac{1}{2}g_s
[A^a,\Phi^3][A^{\bar{a}},\Phi^{\bar{3}}]
\big) \ +\ ... \nonumber
\eeqa
We can perform now the dimensional reduction in the simplest case
of $\Sigma_4=T^4$. Since on this case the profiles of the fields
are constant over the 4-torus, the derivative couplings will not
give any contribution in the four dimensional action, consistently
with our above considerations. Therefore, the integration over
$T^4$ becomes trivial
\beqa
{\mathcal{L}}_{4d}&=&\frac{\mathcal{V}}{(2\pi)^5(\alpha')^2}STr
\left(-\frac{g_s}{18}\left(\frac{1}{4}(S^*_{\bar{1}\bar{2}})^2+
\frac{1}{4}(A^*_{\bar{1}\bar{2}})^2-\frac{1}{2}G^*_{123}S_{33}^*-
\frac{1}{4}S^*_{\bar{2}\bar{2}}S^*_{\bar{1}\bar{1}}\right)\Phi^3\Phi^3+h.c.
-\right. \nonumber \\ & & \left. -\frac{g_s}{18}\left(\vert
G_{\bar{1}\bar{2}\bar{3}}\vert^2+ \frac{1}{4}\vert
S_{\bar{3}\bar{3}}\vert^2 + \frac{1}{4}\sum_{i,j=1,2}(\vert
S_{ij}\vert^2+ \vert A_{ij}\vert^2)\right)\Phi^{\bar{3}}\Phi^3+
\right. \\ \nonumber & & \left.
+\sum_{j,k,p=1,2}\frac{g_s}{3}\epsilon_{3jk}((S_{kp})^*+(A_{kp})^*)
\Phi^3\Phi^{[j}\Phi^{\bar{p}]}+h.c.-\right. \\ & & \left. -
\frac{g_s}{6}\epsilon_{ij3}S_{\bar{3}\bar{3}}\Phi^i\Phi^j\Phi^{\bar{3}}+h.c.
-\frac{2g_s}{3}G_{\bar{1}\bar{2}\bar{3}}\Phi^{\bar{3}}\Phi^{[\bar{1}}
\Phi^{\bar{2}]}+h.c.
-\right. \nonumber \\ & & \left.
-g_s[\Phi^a,\Phi^{(b}][\Phi^{\bar{b})},\Phi^{\bar{a}}]+
\partial_{\mu}\Phi^a\partial_{\mu}\Phi^{\bar{a}}\right) \nonumber
\eeqa
where $\mathcal{V}$ is the volume of $T^4$.

Finally, rescaling all the fields by
$(2\pi)^{5/2}\alpha'{\mathcal{V}}^{-1/2}$ in order to have
canonical kinetic terms, we get the four dimensional action
\beqa {\mathcal{L}}_{4d}&=&Tr
\left(\partial_{\mu}\Phi^m\partial_{\mu}\Phi^{\bar{m}}- \right. \\
\nonumber & &
\left.-\frac{g_s}{18}\left(\frac{1}{4}[(S_{12})^*]^2+\frac{1}{4}[(A_{12})^*]^2
-\frac{1}{2}(G_{\bar{1}\bar{2}\bar{3}})^*(S_{\bar{3}\bar{3}})^*
-\frac{1}{4}(S_{22})^*(S_{11})^*\right)\Phi^3\Phi^3+h.c.-\right.
\\ \nonumber & & \left. -\frac{g_s}{18}\left(\vert
G_{\bar{1}\bar{2}\bar{3}}\vert^2+ \frac{1}{4}\vert
S_{\bar{3}\bar{3}}\vert^2 + \frac{1}{4}\sum_{i,j=1,2}(\vert
S_{ij}\vert^2+ \vert A_{ij}\vert^2)\right)\Phi^{\bar{3}}\Phi^3-
\right. \\ \nonumber & & \left.
-\sum_{k,p=1,2}\frac{g_s^{1/2}g_{YM}}{6}\epsilon_{ijk}
((S_{kp})^*+(A_{kp})^*)\Phi^i\Phi^j\Phi^{\bar{p}}+h.c.+\right. \\
\nonumber & & \left. +
\frac{g_s^{1/2}g_{YM}}{6}\epsilon_{ij3}S_{\bar{3}\bar{3}}
\Phi^i\Phi^j\Phi^{\bar{3}}+h.c.+\frac{g_s^{1/2}g_{YM}}{9}
G_{\bar{1}\bar{2}\bar{3}}\epsilon_{\bar{i}\bar{j}\bar{k}}
\Phi^{\bar{i}}\Phi^{\bar{j}}\Phi^{\bar{k}}+h.c.-\right. \\
\nonumber & & \left.
-g_{YM}^2[\Phi^i,\Phi^{(j}][\Phi^{\bar{j})},\Phi^{\bar{i}}]\right)
\eeqa
with $g_{YM}$ given by
\beq
g_{YM}^2=g_s(2\pi)^5(\alpha')^2{\mathcal{V}}^{-1}
\eeq

\end{document}